\documentclass{aa}  
\usepackage{booktabs}
\usepackage{threeparttable}
\usepackage{graphicx}
\usepackage{txfonts}
\usepackage{lipsum}
\usepackage{subcaption}   
\usepackage{amsmath}
\usepackage{amssymb}
\usepackage{natbib,twoopt}
\usepackage[breaklinks=true]{hyperref} 
\hypersetup{
  colorlinks   = true, 
  urlcolor     = blue, 
  linkcolor    = blue, 
  citecolor   = blue 
}
\usepackage{lscape}             
\usepackage{placeins}           

\usepackage{xspace}

\newcommand{\ha}{H\(\alpha\)\xspace}
\newcommand{\hb}{H\(\beta\)\xspace}
\newcommand{\hg}{H\(\gamma\)\xspace}
\newcommand{\hd}{H\(\delta\)\xspace}
\newcommand{\jwst}{\emph{JWST}\xspace}
\nolinenumbers
\begin{document}
\nolinenumbers
   \title{Evidence of violation of Case~B recombination in Little Red Dots}


   \author{G.~P.~Nikopoulos\inst{1,2}\fnmsep\thanks{georgios.nikopoulos@nbi.ku.dk}
        \and D.~Watson\inst{1,2}\fnmsep\thanks{darach@nbi.ku.dk}
        \and A.~Sneppen\inst{1,2}
        \and V.~Rusakov\inst{1,2,3}
        \and K.~E.~Heintz\inst{1,2,4}
        \and J.~Witstok\inst{1,2}
        \and G.~Brammer\inst{1,2}
        }

   \institute{Cosmic Dawn Center (DAWN), Denmark
            \and Niels Bohr Institute, University of Copenhagen, Jagtvej 155A, DK-2200, Copenhagen N, Denmark
            \and Jodrell Bank Centre for Astrophysics, University of Manchester, Oxford Road, Manchester M13 9PL, UK
            \and Department of Astronomy, University of Geneva, Chemin Pegasi 51, 1290 Versoix, Switzerland}

   \date{Received October XX, 2025}

\abstract{Little Red Dots (LRDs) are a new class of compact extragalactic objects, with a v-shaped optical spectral energy distribution breaking close to the Balmer break wavelength, and broad, typically exponentially-shaped lines. They are believed to be supermassive black holes surrounded by very dense, ionized gas, leading us to explore for any departures from case~B recombination in these systems.}
{In this paper, we examine the ratios of multiple hydrogen Balmer lines: \ha, \hb, \hg, and \hd to determine whether there is indication for deviation from simple case~B recombination processes.}
{We analyze a dozen high-S/N LRDs with \jwst/NIRSpec, measuring Balmer ratios in the seven objects with coverage of at least three lines. We decompose the line ratios into their respective broad and narrow components.}
{Broad line ratios are consistent with Case~B plus severe dust extinction in all objects but one,  RUBIES EGS-49140 at \(z=6.68\), which departs from Case B expectations by more than $5\sigma$. The narrow components are consistent with minimal dust attenuation, while two objects exhibit narrow \ha/\hb $\approx 1.8$. }
{
Such low decrements are observed in highly ionized density bounded nebulae, associated with starburst environments. Nevertheless, both flat decrement cases can be reconciled assuming an unresolved absorption feature. RUBIES EGS-49140, shows a high broad \ha/\hb, but \hg/\ha  and \hd/\ha ratios are lower than expected for extinction-modified case~B, hinting at an unphysically steep dust law. These line ratios may be due to increased optical depth in the Balmer lines, as a direct effect of high density (log$n_e$ > 9) gas surrounding the black hole.  If case~B recombination does hold in most LRDs, they must be moderate-to-heavily dust obscured (\(A_V\simeq1-8\)) while the host-galaxy should be dust-free, suggesting that the extinction in the broad lines is local to the LRD and not due to the general ISM of the host galaxy. 
}

 


   \maketitle

\nolinenumbers
\section{Introduction}

Over the past few years \jwst has  uncovered a growing number of high redshift luminous objects displaying broad recombination lines. Among these broad line systems, a new population of red, compact, point-like sources has emerged. These ``Little Red Dots'' (LRDs, \citealt{Greene2024, Matthee2024}), exhibit blue UV colors and red optical slopes in the rest frame, with the inflection point occurring ubiquitously around the Balmer break \citep{Setton2024}. The observed spectral energy distributions (SEDs) of LRDs could initially be approximately reproduced by a range of models -- from dust-obscured active galactic nuclei (AGN) with minimal stellar mass, to massive stellar populations, or a combination of both \citep{Killi2024,Rinaldi2024,Wang2024b} -- but now, more extreme Balmer breaks in the SEDs clearly rule out a stellar origin or standard AGN scenario \citep{Naidu2025,deGraaff2025Cliff}.
With the nature of LRDs still under debate, the prevailing consensus suggests that they host supermassive black holes (SMBHs), based on the broad width and high luminosities of their Balmer lines and their compact spatial sizes \citep{Harikane2023, Maiolino2024, Rusakov2025}.

The interpretation that these objects are AGN has faced several challenges. Firstly, recent work reports a lack of x-ray emission, even with non-detections in deep Chandra imaging, \citep{Akins2024, Ananna2024, Kokubo2024, Maiolino2024Xray, Yue2024}, as well as an absence of radio emission \citep{Akins2024, Gloudemans2025, Mazzolari2024, Setton2025}. Secondly, their rest-frame mid- and far-IR SEDs are flat, implying that their red colours are not primarily due to extreme dust extinction \citep{PerezGonzalez2024}, combined with a lack of hot, dusty torus emission \citep{Akins2024, Williams2024, Setton2025}. As mentioned above, many objects show Balmer breaks with varying strengths \citep{Setton2024, deGraaff2025Cliff, Ji2025} with \cite{Naidu2025} reporting the strongest Balmer break to date, indicating a possible old stellar component in their SED \citep{Wang2024b}, or, more likely, absorption by dense gas surrounding the broad line region (BLR) \citep{Rusakov2025,Naidu2025,Inayoshi2025,Kido2025}. Furthermore, if the Balmer line widths are dominated by Doppler broadening, then the inferred black holes are too massive compared to the local $M_{\mathrm{BH}}-M_{\mathrm{star}}$ relation given their host galaxy stellar mass \citep{Maiolino2024}, and given the high redshift of these objects $z \gtrsim 4$, this may require heavy black hole seeds masses \citep{Trinca2023}. Recent work by \citet{Rusakov2025} reconciles some of the above tensions by showing that the dominant permitted-line broadening is due to electron scattering of photons in a very dense cocoon of ionized gas surrounding the black hole \citep[see also e.g.][]{Chang2025,D'Eugenio2025deviation, Torralba2025}. This results in 1--2\,dex lower black hole masses, and provides a natural explanation for the x-ray weakness of these objects, as they would be high accretion rate objects with strong Compton cooling of the corona producing the hard x-rays.

The V-shaped continua of LRDs pose yet another puzzle; the red optical slopes suggest significant dust-attenuation ($A_V \simeq 2-4$), while their blue UV spectra imply $A_V \simeq 0$ \citep{Killi2024,Casey2024}. Recent \jwst studies have investigated the dust content in LRDs in an attempt to resolve this. \citet{Rinaldi2024} use SED modeling on a sample of LRDs to retrieve an average $A_V \simeq 1.2$ under pure stellar assumptions, while using an AGN template yields $A_V \simeq 2.7$. Spectroscopic studies of the Balmer decrement (\ha/\hb) in LRDs further reveal that narrow lines are consistent with no extinction (\ha/\hb $\simeq 2.5$), while for the broad components, \ha/\hb $\gtrsim 8.9$ ($A_V \gtrsim 3.6$) on average, indicating severe dust extinction in the broad line region \citep{Brooks2025}. Also, \citet{Killi2024} find similar results, by estimating the $A_V$ both from continuum fits and the narrow and broad Balmer decrements. 
\begin{table*}[t!]
\centering
\renewcommand{\arraystretch}{1.4}
\begin{tabular}{@{}ccccc@{}}
\toprule
\multicolumn{1}{c}{ID} & \multicolumn{1}{c}{Survey-Field-MSAID} & $\alpha$ ($^{h}$:$^{m}$:$^{s}$) & $\delta$ ($^{o}$:$'$:$"$) & Redshift \\ \midrule
A                      & JADES-GN-68797                         & 12:36:54.993                    & 62:08:46.283              & 5.0405   \\
C                      & JADES-GN-73488                         & 12:36:47.375                    & 62:10:38.039              & 4.1327   \\
E                      & RUBIES-EGS-49140                       & 14:19:34.139                    & 52:52:38.675              & 6.6847   \\
F                      & CEERS-EGS-1244                         & 14:20:57.757                    & 53:02:09.748              & 4.4771   \\
I                      & JADES-GN-53501                         & 12:37:10.814                    & 62:11:36.860              & 3.4294   \\
J                      & JADES-GN-38147                         & 12:37:04.962                    & 62:08:54.307              & 5.8694   \\
K                      & RUBIES-EGS-50052                       & 14:19:17.629                    & 52:49:48.996              & 5.2392   \\ \bottomrule
\end{tabular}
\caption{The ID, Survey-ID, right ascension ($\alpha$), declination ($\delta$) and  redshift of the objects used in this work. Objects A, F, I and J are shown as a extracted from the DJA. References to the rest of the objects are given: C; \cite{Maiolino2024}, E; \cite{Kocevski2025, Wang2024b}, K; \cite{Harikane2023}. The typical uncertainties in the redshift are about $3 \times 10^{-4}$, while $\alpha$ and $\delta$ measurements are J2000.0.}
\label{table1-lrds}
\end{table*}
However, caution must be used when using the Balmer decrement to infer the dust properties of LRDs. In very high electron density conditions (\(\log{n_e} > 8\), \citealt{Netzer1975, Drake1980}), Case~B may not be a valid assumption: self-absorption of Balmer photons, collisional excitation to $n>2$ and collisional de-excitation, render the Balmer lines optically thick. Both dust extinction and such deviations from Case~B may modify the Balmer decrement in similar ways \citep{Netzer1975, Drake1980}. This degeneracy can be lifted by examining higher-order Balmer lines (e.g.\ \hg, \hd): simple dust extinction reduces the \hg/\ha and \hd/\ha ratios in a manner that is consistent with the adopted extinction curve and inferred $A_V$, whereas the effects of high density can alter each ratio in a different way, depending on the local nebular conditions. As these changes can be accurately predicted with radiative-transfer calculations \citep[see][for a discussion of the above in the context of LRDs]{Chang2025}, the observed Balmer ratios can in principle be used to constrain the density of the emitting gas, and to some degree, the electron temperature ($T_e$). Thus, in the context of high~$z$ AGN and LRDs, by analyzing multiple Balmer line ratios and treating the broad and narrow components separately, one can (i) estimate the various $A_V$ values for objects consistent with Case~B, determining whether the red continuum is primarily due to a dust-reddened AGN with the blue UV originating from star formation (or vice versa), and (ii) when deviations from Case~B are present, use the pattern of line ratios to estimate physical conditions in the BLR and NLR.

In this work, we use medium and high resolution spectra to decompose the \ha, \hb, \hg and \hd lines of a sample of high SNR LRDs into their narrow and broad components. We examine potential deviations from Case~B recombination by computing the flux ratios for every component. We introduce the spectroscopic sample and describe our methodology in Section~\ref{sec-observations-and-methods}. The measured flux ratios, the inferred $A_V$ values, as well any departures from Case~B are presented in Section~\ref{sec-results}. We discuss the implications of these findings on the nature of LRDs in Section~\ref{sec-discussion}, and conclude with a summary in Section~\ref{sec-summary}. Upper limits are quoted at the 95\% confidence limit. Detections are claimed at \(3\sigma\) or greater significance.

\section{Observations and Methods}
\label{sec-observations-and-methods}
\subsection{Observations}
We build on the work of \citet{Rusakov2025} by analyzing a subset of the LRDs included in their sample. Their dataset consists of broad-line \jwst galaxies, with \ha linewidths greater than $1000\,\mathrm{km}\,\mathrm{s}^{-1}$,  median SNR $>$ 5 per 10\AA\ for the continuum-subtracted region $\pm\;2000\,\mathrm{km}\,\mathrm{s}^{-1}$ around the \ha line, available in medium or high resolution grating spectra. They also include a stack of objects with lower SNR (1 < SNR/10$\,$\AA$\,$< 5) spectra. From this parent sample, we select objects with medium- or high-resolution spectral coverage of the \ha, \hb, and \hg lines, while the \hd line is also used when available. The data are publicly available from several \jwst observing programs using the NIRSpec spectrograph \citep{Jakobsen2022} with Program IDs: 1181 (JADES; \citealt{Eisenstein2023}), 4233 (RUBIES; \citealt{deGraaff2025RUBIES}) and 1345 (CEERS; \citealt{Finkelstein2023}). We use version 4 (`v4') of the reduced spectra, retrieved from the DAWN \jwst Archive (DJA\footnote{DOI: 10.5281/zenodo.8319596}, \citealt{deGraaff2025RUBIES,Heintz2025, Pollock2025}). Detailed information on our sample is provided in Table~\ref{table1-lrds}, while the corresponding SEDs are shown in the Appendix (Figure~\ref{fig:appendix-SEDs}).

\subsection{Methods}

\begin{figure}[h]
    \centering
    \includegraphics[width=0.8\linewidth]{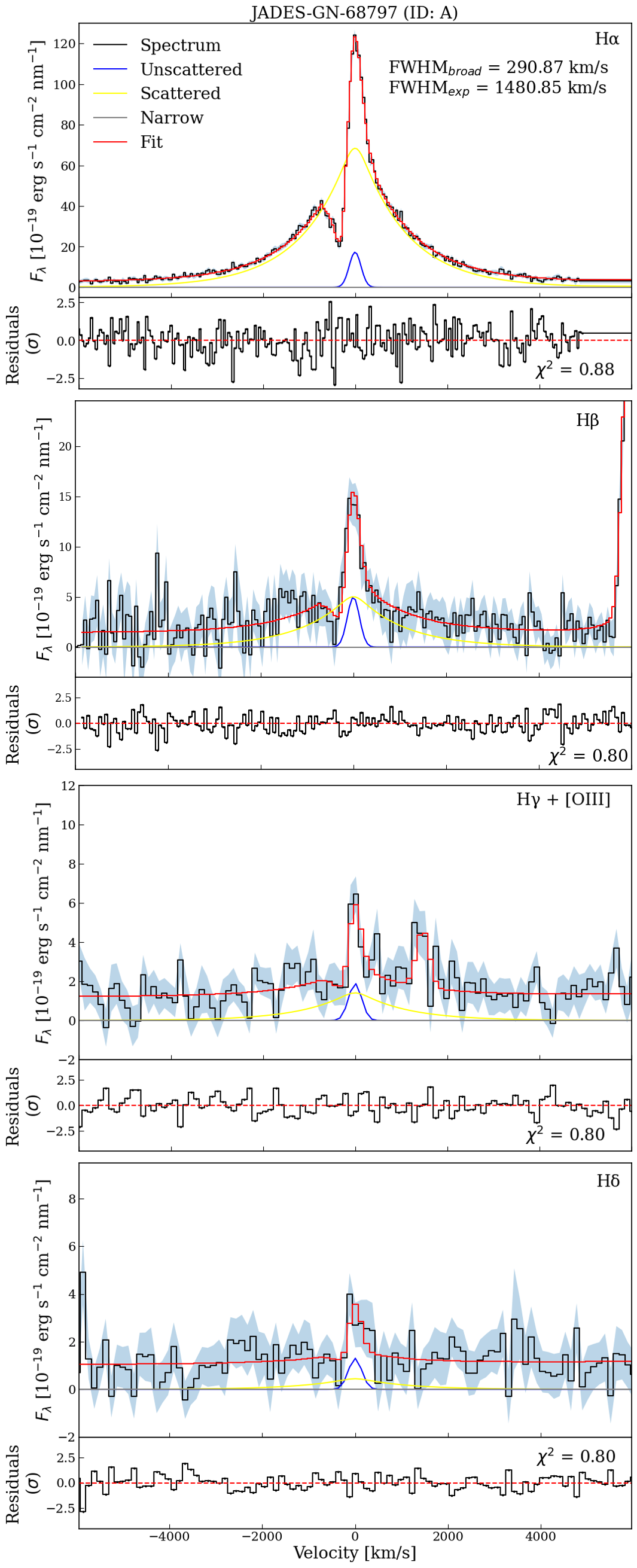}
    \caption{The best-fit solutions for the \ha, \hb, \hg and \hd of JADES-GN-68797 (ID A), as noted in their respective panels. The separate components that make up the line profile are also shown. We note that there is no narrow emission detected in JADES-GN-68797.}
    \label{fig-object-a}
\end{figure}

A key property of many LRDs is the extended wings found in the \ha emission lines. In our sample we also find similarly broad \hb and \hg emission in various objects (see for example Object~A; Figure~\ref{fig-object-a}). 
We assume that the Balmer emission in these LRDs is composed of a narrow intrinsic line core and exponential wings, as a result of electron scattering through a Compton-thick medium \citep{Rusakov2025}. The above framework represents a central engine (for example, an accreting supermassive black hole), embedded in a cocoon of highly dense, ionized gas \citep{Rusakov2025,Naidu2025,deGraaff2025Cliff,Inayoshi2025}. Following the work of \citet{Rusakov2025}, our \textit{fiducial} model consists of three components:
\begin{itemize}
    \item \emph{A narrow component}: a Gaussian emission line of width $\sigma_{nr}$, corresponding to emission coming either from the host galaxy, or a narrow line region. 
    \item \emph{An unscattered component}: a Gaussian emission line with a width $\sigma_g$, describing the light escaping the dense medium without being scattered.
    \item \emph{A scattered component}: a convolution of the above non-scattered gaussian with an exponential of width $\sigma_{exp}$, to account for electron scattering of photons.
\end{itemize}
    
\noindent The sum of the unscattered and scattered components, (taking into account possible absorption) is hereafter referred to as the \textit{broad} component. 

The \textit{fiducial} model is used to fit the \ha lines in our sample. We tie the narrow \ha width ($\sigma_{nr}$) and redshift to those of the [O\,{\sc iii}]4959,5007\AA\ doublet in velocity space, by introducing a flat 50\,km\,s\(^{-1}\) prior around the latter. To fit the broad \ha components, in Object~A, we also use a P~Cygni profile \citep[e.g.\ see][]{Sneppen2023}, while in object~E a Gaussian absorption component is used. Adding the [N\,\textsc{ii}] doublet in our model does not significantly improve any of the \ha fits and so its use is omitted. All widths are corrected for the instrumental resolution of the gratings. The actual resolution of all gratings is taken to be 1.7 times higher than the nominal values cited in the NIRSpec documentation, as suggested by instrument modelling of point sources in the literature \citep{deGraaff2024}.

After retrieving the \ha profile, we fit \hb, \hg and \hd with the same widths, redshift and (where applicable) P~Cygni parameters \citep[see][]{Rusakov2025} as the respective \ha lines, but with different amplitudes. We choose to treat the non-scattered and scattered components as one component with a single amplitude, in order to avoid overfitting in cases where the broad components are weak or not detected altogether. We determine that the above provides a robust fit to observations in all cases, with the exception of Object~E; the \hb line of object E requires an absorption profile with a different width, systemic velocity and amplitude than the \ha absorption profile. In general, there are good reasons to believe that the profiles of the Balmer lines might be the same or different due to radiative transfer effects, depending on how extreme the column density is, but in most of these cases, the fits were good adopting this technique. 

In  this work, we use \texttt{dynesty} v.2.1.5 \citep{Speagle2020,sergey_koposov_2024_12537467} with static nested sampling \citep{Skilling2004, Skilling2006}, adopting $75n$ live  points (walkers), where $n$ is the number of free parameters. To efficiently explore complex posterior landscapes, we use random-walk sampling \citep{Skilling2004} and multi-ellipsoidal bounding \citep{Feroz2009} around the set of live points, to account for possible degeneracies and multimodality. We note that \texttt{dynesty} is not a traditional optimizer, as it does not explicitly search for a global minimum. The sampling is terminated once the change in the log-evidence is lower than a user-specified threshold. Thus, after obtaining the highest likelihood posterior distributions for all free parameters with \texttt{dynesty}, we use them as initial conditions for a Levenberg-Marquardt minimization algorithm via \texttt{lmfit} v.1.3.2 \citep{lmfit-newville2024}, from which we derive our global best-fit solution. We provide the best-fit solutions for the \ha, \hb, \hg, and \hd lines of JADES-GN-68797 (Object~A) in \ref{fig-object-a}. We note that there is no narrow emission detected in this object. As a result, for JADES-GN-68797 we treat the non-scattered and the scattered components separately (using different amplitudes per component) when fitting \hb, \hg and \hd. The best-fit solutions for the rest of the objects in our sample,  are presented in the Appendix.
After repeating the above procedure for each Balmer line (\ha, \hb, \hg, and \hd), we use the posterior distributions from \texttt{dynesty} to compute flux distributions for both the narrow and broad components of the lines. The best-fit flux for each component corresponds to the flux derived from integrating our model evaluated at the global best-fit parameters, while uncertainties are taken as the 16th and 84th percentiles of the derived flux distribution. We consider the component of a line to be detected, if its flux is more than 3$\sigma$ above zero.

We then propagate these flux posteriors to derive distributions of the flux ratios of interest (e.g.\ \ha/\hb). The best-fit value of a line ratio corresponds to the ratio of the respective best-fit fluxes, while the uncertainties are derived from the 16th and 84th percentiles of the distribution of ratios. Throughout this paper, in case the component of a line in the numerator is not detected, we report the 2$\sigma$ upper limit on the ratio.

\section{Results}
\label{sec-results}

\begin{figure}
    \includegraphics[width=\columnwidth]{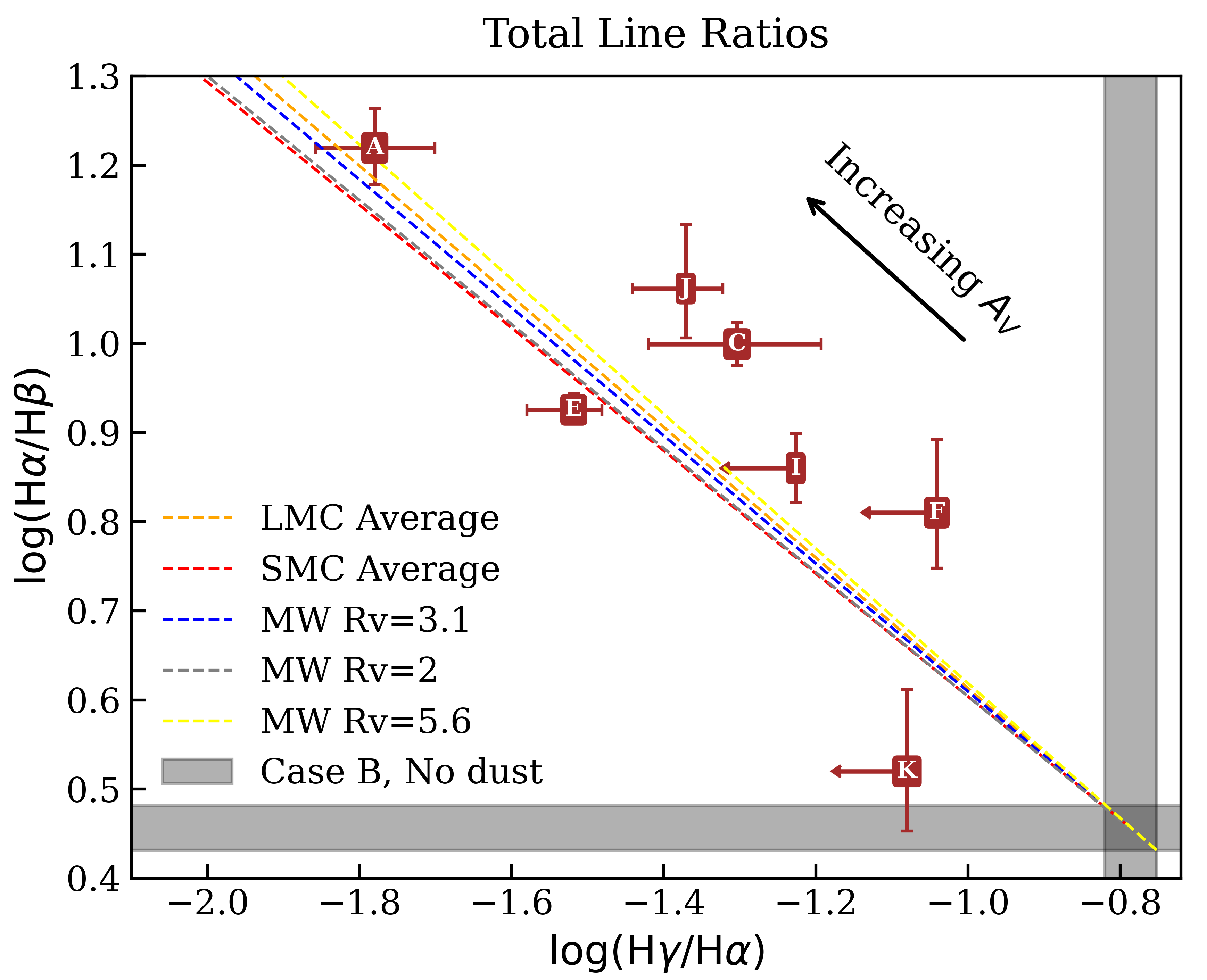}
       \caption{Balmer line ratio diagram showing total line \ha/\hb vs. \hg/\ha. The black shaded regions mark the intrinsic Case~B ratios across a range of reasonable values,
       while the colored dashed lines show the LMC \citep{lmc}, SMC \citep{Gordon2024-SMC} and Milky Way \citep{mw} extinction laws, derived using the \texttt{dust\_extinction} package \citep{Gordon2024-dust_ext}. The arrow indicates the direction of increasing dust attenuation.}
          \label{fig:total_line_ratios}
\end{figure}

We hypothesize that the broad- and narrow-line components could represent emission from distinct regions and could possibly be associated with different gas densities, temperatures, and extinctions. In such a framework, the flux ratios of the total line (the sum of both broad and narrow) gives a potentially misleading picture, yielding a heterogeneous landscape of line-ratios. In particular, we find the total line ratios for the various objects sometimes do not yield line ratios that can be made consistent with Case~B recombination even with arbitrary extinction for any reasonable extinction law (see Figure~\ref{fig:total_line_ratios}). 
\begin{figure*}
    \includegraphics[width=0.9\textwidth]{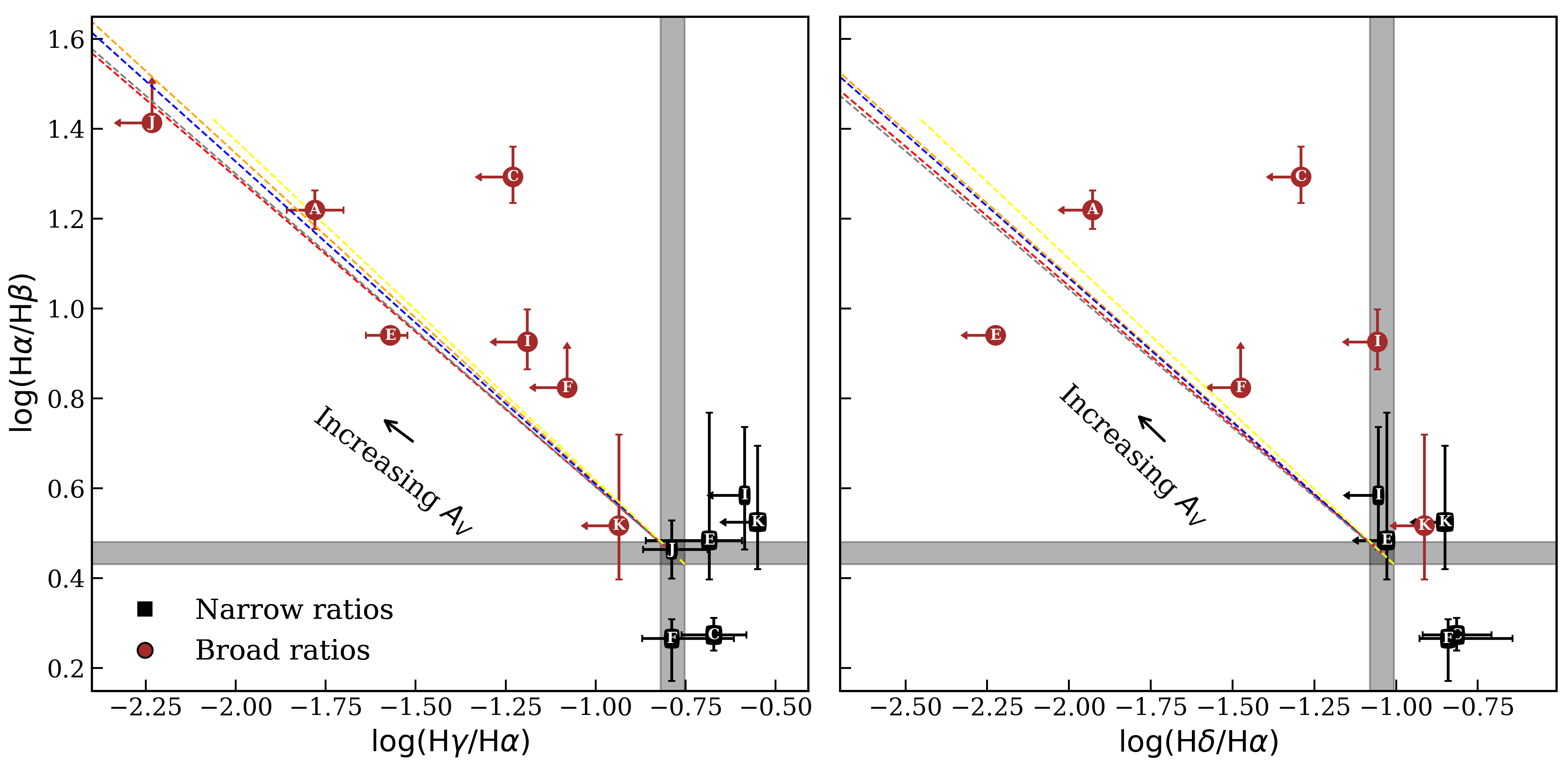}
       \caption{Balmer line ratio diagrams showing the narrow and broad \ha/\hb vs.\ \hg/\ha (\emph{left}) and \ha/\hb vs.\ \hd/\ha (\emph{right}) ratios. Broad ratios are shown as brown circles and narrow ratios as black squares. The black shaded regions mark the intrinsic Case~B ratios across a range of reasonable values. The colored dashed lines show the LMC \citep{lmc}, SMC \citep{Gordon2024-SMC}, and Milky Way \citep{mw} extinction laws, derived using the \texttt{dust\_extinction} package \citep{Gordon2024-dust_ext}. The arrows indicate the direction of increasing dust attenuation.}
          \label{fig-ratios-main}
\end{figure*}

Nevertheless, when decomposed, the line ratios of the narrow and broad components present distinct patterns, most of which \emph{are} consistent with Case~B plus dust extinction. This result shows that it is important to decompose the lines into their respective physical components. Figure~\ref{fig-ratios-main} shows the ratios \ha/\hb, \hg/\ha and \hd/\ha observed for both the broad and narrow lines. Comparing this with the expected ratio for Case~B with different extinction curves \citep[SMC, LMC, MW,][]{Gordon2024-dust_ext} as a function of $A_V$ reveals several key results.

As clearly seen from \ha/\hb, the narrow component is generally consistent with no extinction, whereas the broad component requires a substantial \(A_V\). Specifically, the Balmer decrement (\ha/\hb) is on average 2.81 and 11.3 for the narrow and broad components, respectively. For a Case~B coefficient with \(2.7<\)\ha/\hb\(<3.03\) ($n_e = 500\,\,\rm{cm^{-3}}$, $5000\,\rm{K}\, < \rm{T}< 30000 \,\rm{K}$), this corresponds to $A_{V,~\mathrm{narrow}} \sim 0$ and $A_{V,~\mathrm{broad}} = 4.1\pm0.2$. 

The increase in the \ha/\hb ratio can be facilitated by both dust extinction and high density effects that can make the Balmer lines optically thick \citep[e.g.\ self-absorption conditions,][]{Netzer1975, Drake1980}. Examining another line-ratio can break this degeneracy, so we extend our analysis to \hg/\ha in Figure~\ref{fig-ratios-main} (left panel). While several broad lines only have tentative detections or non-detections in \hg, both objects~A and E include clear detections. While A remains broadly consistent with Case~B, we note that E shows a mild 1.8$\sigma$ deviation from Case~B, assuming (\ha/\hb)$_{\mathrm{Case~B}} = 2.7$.

\begin{figure*}
    \centering
    \includegraphics[width=0.9\textwidth]{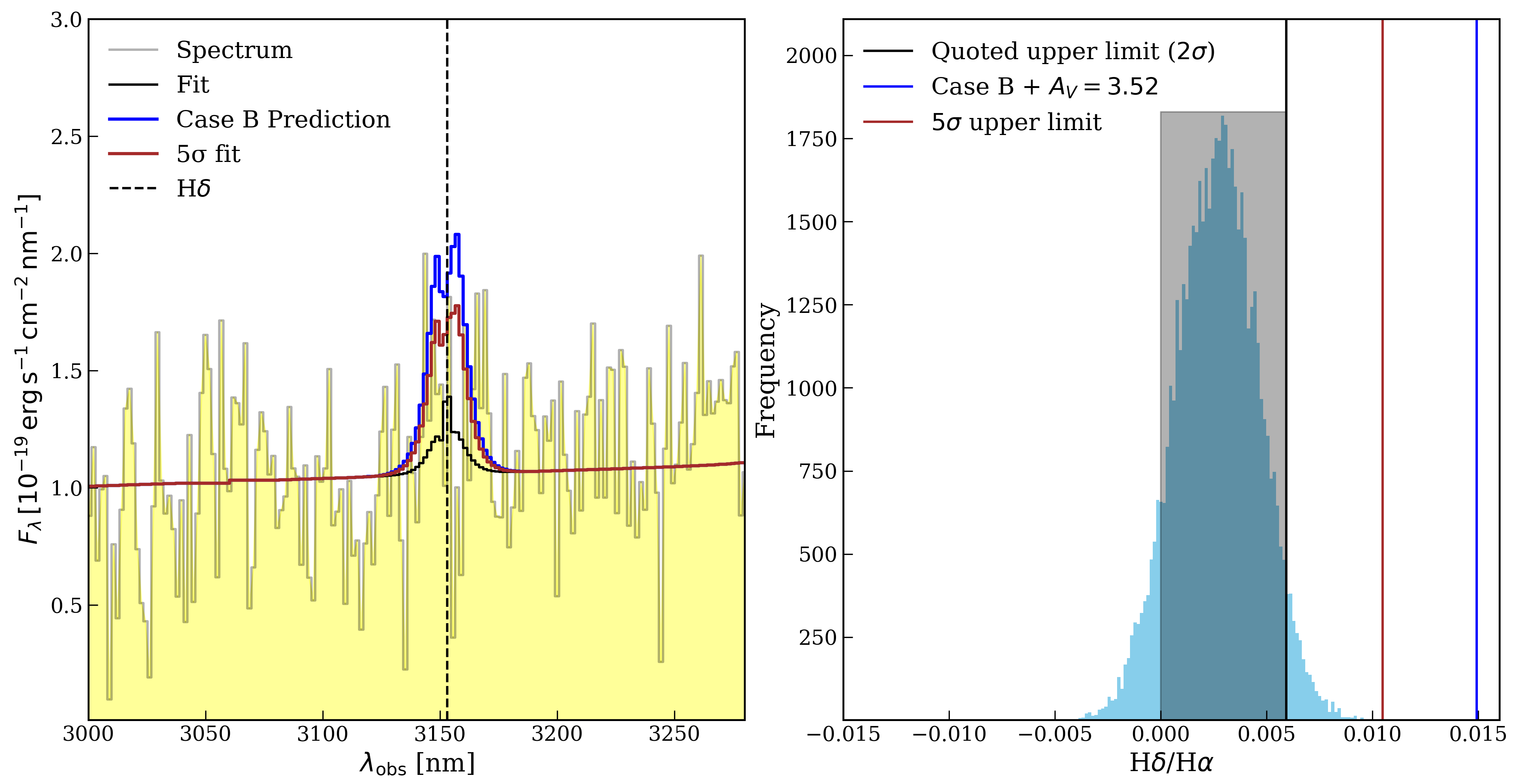}
       \caption{\emph{Left:} The \hd line of Object~E (RUBIES-EGS-49140). The data are shown with a yellow fill and the best fit with a black line, the modeled \hd line profile based on the 5$\sigma$ upper limit of its flux as a brown line, and the predicted \hd line profile under Case~B recombination assumptions as a blue line. The best fit solution yields $\chi^2_\nu = 1.08$, while the $5\sigma$ and the Case~B predicted solutions yield $\chi^2_\nu = 1.15$ and $\chi^2_\nu = 1.27$ respectively. \emph{Right:} The \hd/\ha flux ratio posterior distribution. The black and brown lines represents the 2 and 5$\sigma$ upper limits respectively. The blue line indicates the expected line ratio under Case~B recombination and the $A_V = 3.52$ derived from the \ha/\hb ratio. Assuming a steeper curve, or a higher intrinsic \ha/\hb ratio increases the expected \hd/\ha ratio, making the deviation even more significant.}
          \label{fig-hdha-objectE}
    \end{figure*}

To test whether this inferred mild tension with Case~B for object~E is meaningful, we can use the next line-ratio in the Balmer series as further corroboration, i.e.\ \hd/\ha (see Figure~\ref{fig-ratios-main}, right panel). The broad \hd component is not detected for any object, but for object~E even the 2$\sigma$ upper-limit of (\hd/\ha)$\,<\,0.006$ is substantially below the range predicted for any dust extinction law and the broadest range of Case~B recombination ratios. In particular, as shown in Figure~\ref{fig-hdha-objectE}, this line ratio deviates more than $5\sigma$ from the expected Case~B recombination ratio, assuming (\ha/\hb)$_{\mathrm{Case~B}} = 2.7$. 
Overall the broad line ratios of all objects in our sample apart from Object~E can be consistent with Case~B recombination with extinction. Nevertheless, we caution that the \hg broad component is undetected in all objects except A and E, while we place only upper limits for the \hd broad component in all objects. Observations with enough exposure time to detect the broad component of higher order Balmer lines are required to ascertain whether deviation from Case~B recombination is a common occurrence in LRDs. 
\begin{table*}[t]

\begin{threeparttable}
\centering
\renewcommand{\arraystretch}{1.4}
\begin{tabular}{@{}cccccccccc@{}}
\cmidrule(l){3-10}
\multicolumn{2}{c}{}    & \multicolumn{2}{c}{\ha/\hb} & \multicolumn{2}{c}{\hg/\ha} & \multicolumn{2}{c}{\hd/\ha} & \multicolumn{2}{c}{$A_V$}                        \\ \cmidrule(l){3-10} 
ID & Survey-Field-MSAID & Narrow                      & Broad                       & Narrow                     & Broad                        & Narrow                           & Broad                  & Narrow                  & Broad            \\ \midrule
A \tnote{1}  & JADES-GN-68797     & -                           & $16.57^{+1.66}_{-1.62}$     & -                          & $0.017^{+0.003}_{-0.003}$    & -                                & $ < 0.012$             & -                       & $5.14^{+0.24}_{-0.21}$ \\
C \tnote{2}  & JADES-GN-73488     & $1.88^{+0.16}_{-0.15}$      & $19.64^{+3.16}_{-2.61}$     & $0.21^{+0.04}_{-0.04}$     & $ < 0.059$                   & $0.15^{+0.04}_{-0.04}$           & $ < 0.052$             & $-0.96^{+0.20}_{-0.20}$ & $5.82^{+0.50}_{-0.39}$ \\
E \tnote{2} & RUBIES-EGS-49140   & $3.05^{+1.97}_{-0.61}$      & $8.72^{+0.37}_{-0.39}$      & $0.21^{+0.04}_{-0.08}$     & $0.027^{+0.003}_{-0.004}$    & $ < 0.094$                       & $ < 0.006$             & $0.21^{+0.33}_{-0.27}$  & -                      \\
F \tnote{2}  & CEERS-EGS-1244     & $1.84^{+0.19}_{-0.40}$      & $> 6.83$ \& $ < 0.083$      & $0.16^{+0.06}_{-0.03}$     & $ < 0.032$                   & $0.14^{+0.07}_{-0.03}$           & $> 2.45$               & $-0.15^{+0.44}_{-0.29}$ & $> 2.45$               \\
I  & JADES-GN-53501     & $3.84^{+1.37}_{-1.08}$      & $8.44^{+1.43}_{-1.21}$      & $< 0.25$                   & $ < 0.066$                   & $ <  0.09$                       & $ < 0.088$             & $1.28^{+2.13}_{-0.81}$  & $3.28^{+0.56}_{-0.41}$ \\
J  & JADES-GN-38147     & $2.91^{+0.45}_{-0.42}$      & $ > 25.22$                  & $0.16^{+0.04}_{-0.03}$     & $ < 0.006$                   & -                                & -                      & $0.01^{+0.40}_{-0.23}$  & $ > 7.00 $             \\
K  & RUBIES-EGS-50052   & $3.35^{+1.26}_{-0.78}$      & $3.29^{+1.54}_{-0.91}$      & $< 0.28$                   & $ < 0.012$                   & $ < 0.14$                        & \{$ < 0.12$\}          & $0.63^{+1.38}_{-0.78}$  & $1.72^{+2.20}_{-0.87}$ \\ \bottomrule
\end{tabular}
\begin{tablenotes}
\footnotesize
\item[1] JADES-GN-68797 does not have a detected narrow component in any line examined.
\item[2] These objects deviate from Case~B recombination (see Section 
\ref{sec-results}). They are the only objects whose extinction curve fits yield $\chi_{\nu}^2 >1$.
\end{tablenotes}
\caption{Narrow and broad line ratios as well as the corresponding narrow and broad $A_V$ values estimated from simultaneously fitting all available line ratios.}
\label{tab:broad-avs}

\end{threeparttable}

\end{table*}

\subsection{Deviations from Case~B in the Narrow Line Ratios}
\label{sec-res-dev-in-narrow}

\begin{figure}[h!]
    \centering
    \includegraphics[width=\columnwidth]{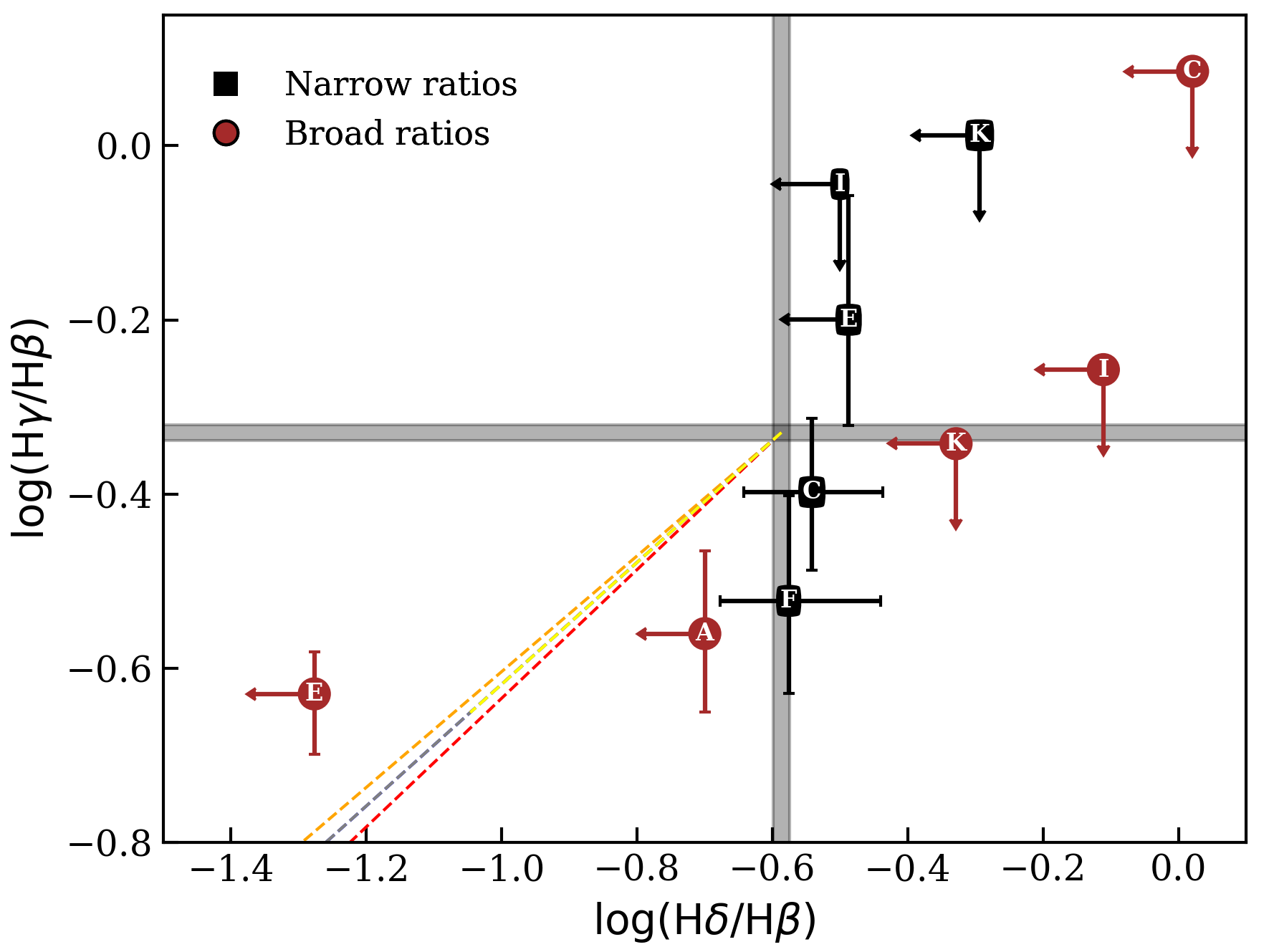}
       \caption{Balmer line ratio diagrams showing the narrow and broad \hg/\hb vs. \hd/\hb ratios for our sources. Broad ratios are shown as brown circles and narrow ratios as black squares. The black shaded regions mark the intrinsic Case~B ratios across a range of reasonable values. The colored dashed lines show the LMC \citep{lmc}, SMC \citep{Gordon2024-SMC}, and Milky Way \citep{mw} extinction laws, derived using the \texttt{dust\_extinction} package \citep{Gordon2024-dust_ext}. The arrows indicate the direction of increasing dust attenuation.}
          \label{fig-hghb-vs-hdhb}
\end{figure}

We highlight objects~C (JADES-GN-73488) and F (CEERS-EGS-1244) for their unusually flat Balmer decrements. In both objects, the narrow component shows a \ha/\hb ratio well below the lowest expected Case~B ratio of $\simeq 2.7$. Specifically, Object~C has a narrow \ha/\hb $\simeq 1.88^{+0.16}_{-0.15}$, a $5\sigma$ deviation from Case~B (\ha/\hb = 2.7), implying an unphysical $-1.44 < A_V < -1.09$, for the Case~B Balmer decrement range \(3.03 > \)\ha/\hb\(> 2.7\) explored in this work. The higher-order ratios are \hg/\ha $= 0.21\pm0.04$ and \hd/\ha $= 0.15^{+0.04}_{-0.04}$, fairly consistent with Case~B. Object~F shows a very similar pattern; its narrow Balmer decrement, \ha/\hb $\simeq 1.84^{+0.19}_{-0.40}$ is  $4.6\sigma$ deviant from Case~B ($-1.49 < A_V < -1.15$), while \hg/\ha $= 0.163^{+0.76}_{-0.03}$ and \hd/\ha $= 0.144^{+0.07}_{-0.03}$, are consistent with Case~B with no dust extinction within about 90\% confidence. In other words, for both objects, either the \ha line is too weak by about 50\% or the \hb line is too strong by about 50\% compared to the other Balmer lines to be consistent with Case~B recombination ratios. The above can be investigated further by examining other combinations of line ratios in our sample. To this end, we present the \hg/\hb vs.\ \hd/\hb ratios in Figure~\ref{fig-hghb-vs-hdhb}. 

The constraining power of most line ratios in Figure~\ref{fig-hghb-vs-hdhb} is limited due to the fact that \hg and \hd are either weakly detected, or undetected. Nevertheless, looking at the narrow line ratios for Objects~C and F we can deduce the following:  the narrow \hg/\hb and \hd/\hb ratios are broadly consistent with Case~B and no dust extinction. This would imply that the narrow \ha flux is lower than expected, either because of intrinsic suppression or an underestimation in its measurement. We revisit this point in more detail in Section~\ref{sec-narrow-objects}.

By contrast, the broad line components of both objects can be consistent with Case~B ratios with heavy extinction. Specifically, the inferred \ha/\hb = $19.64^{+3.16}_{-2.62}$ (Object~C) and \ha/\hb$ > 6.83$ (Object~F), while their \hg and \hd are undetected; we place upper limits for both ratios, of \hg/\ha \(< 0.059\) (\hg/\ha \(<  0.083\)) and \hd/\ha \(<  0.052\) (\hd/\ha \(<  0.032\)) for object~C (object~F). The Balmer decrements imply a range of  $5.96 > A_V > 5.61 $ and $A_V > 2.45$ for Objects~C and F respectively.  

This dichotomy of relatively dust-free (and even flatter than Case~B) narrow line ratios, along with severely reddened broad emission, reinforces the interpretation that the narrow and broad lines arise from spatially and physically distinct regions.


\subsection{Deriving the extinction in our sample}
\label{sec-res-extinctioncurves}
\begin{figure}[h]
    \centering
    \includegraphics[width=\hsize]{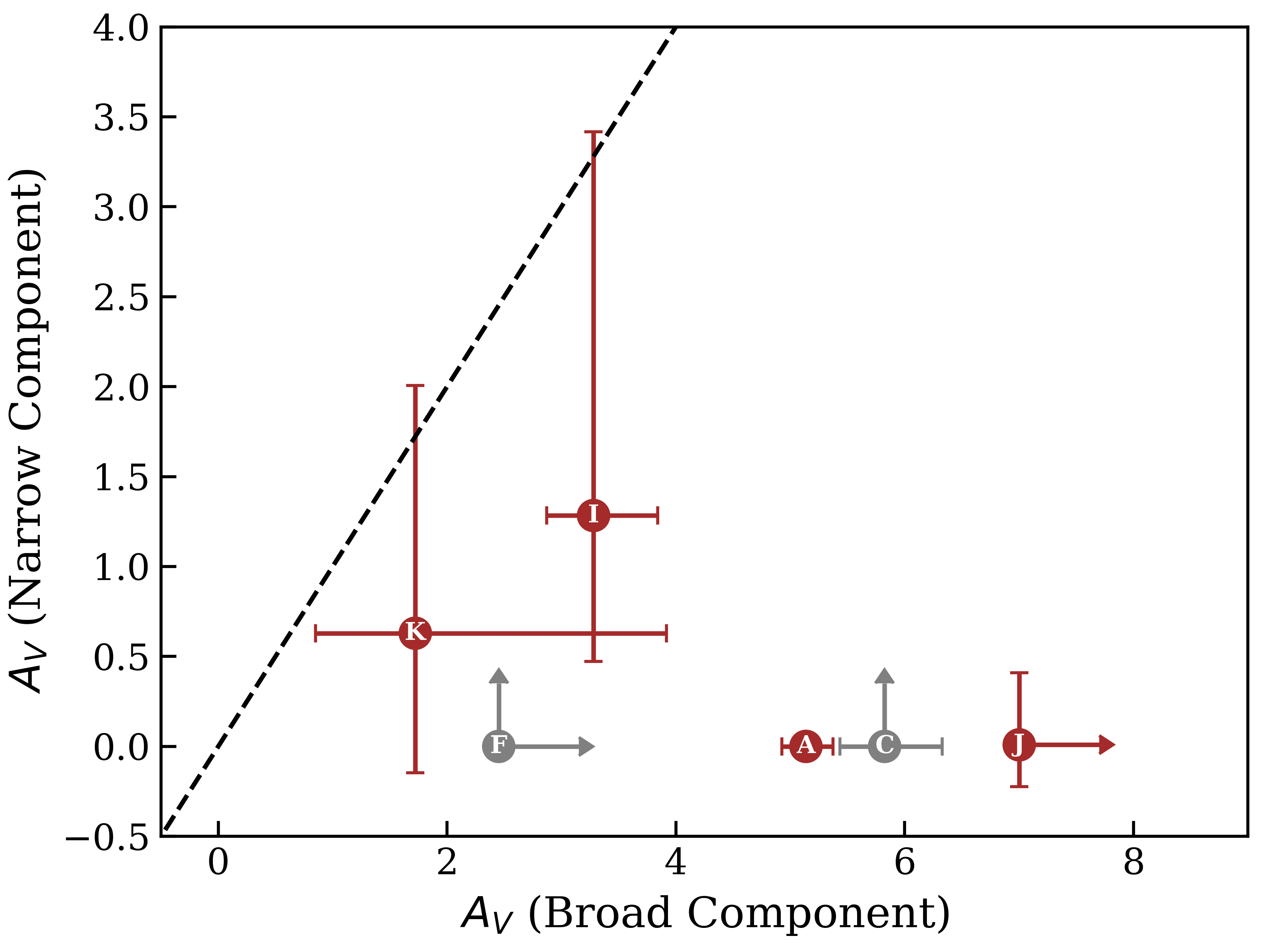}
       \caption{The nebular attenuation inferred from the narrow and broad line ratios. Objects deviating from Case~B in the narrow component are plotted as $A_V > 0$ lower limits and are denoted in gray. The black dashed line represents the $A_V$ (narrow) = $A_V$ (broad) case. We adopt an SMC extinction curve \citep{Gordon2024-SMC}.}
          \label{fig-av-estimation}
\end{figure}

We finally estimate the different $A_V$ values implied by the narrow and broad line ratios calculated in this work. These calculations are done based on all three line ratios (except for Object~J, for which there is no medium resolution coverage of the \hd line). In this way, we derive a more robust measurement of $A_V$ compared to one derived solely from the \ha/\hb ratio. To this end, we assume an intrinsic \ha/\hb ratio of $2.86$ and we use the \texttt{dust\_extinction} package v.~1.5 \citep{Gordon2024-dust_ext}, adopting a Small Magellanic Cloud average extinction curve ($R_V = 3.02$, \citealt{Gordon2024-SMC}). The object IDs, line ratios and estimated $A_V$ values using the above method are presented in Table~\ref{tab:broad-avs}. As discussed in the previous subsection, the narrow component is on average consistent with little-to-no dust, while the broad component is severely reddened. This discrepancy between the narrow and broad $A_V$ is highlighted in Figure~\ref{fig-av-estimation}. 

\section{Discussion}
\label{sec-discussion}
In this work, we examine the line ratios between \ha, \hb, \hg and \hd (where there is spectral coverage) and we find that most objects in our high-$z$ AGN sample can be made consistent with Case~B recombination assumptions. Nevertheless, we have found two objects deviating in the narrow component, and one broad component (one of two objects for which the broad \hg is detected) is inconsistent with Case~B. In this section we examine these three objects in sections~\ref{sec-narrow-objects} and \ref{sec-objectE}, while in sections~\ref{sec-agn-vs-sf} and \ref{sec-ir-problem} we discuss the implications of the results of this work.

\subsection{Flat narrow Balmer line ratios}
\label{sec-narrow-objects}
JADES-GN-73488 (Object~C) and CEERS-EGS-1244 (Object~F) show particularly peculiar narrow-line behavior, as shown in section~\ref{sec-res-dev-in-narrow}; both showing narrow \ha/\hb $< 2$ with a deviation greater than $4\sigma$ from Case~B recombination. Meanwhile, their narrow \hg/\ha  and \hd/\ha ratios are consistent with Case~B, within about $\sim 1.5\sigma$. 

Because such results are surprising and could in principle be driven by modeling choices, we explicitly tested the robustness of the narrow-component measurement by refitting each object with several alternative parameterisations. We tested 1) a case where there is no escaping fraction of unscattered light, i.e.\ removing the unscattered Gaussian component, 2) a case where no narrow line/host emission is detectable, i.e.\ removing the narrow Gaussian component, and 3) a model which assumes the broad line-shape of these objects is not exponential, with their broad component described by two Gaussians. The resulting fits of all three scenarios are substantially worse than the fit presented in the appendix for these objects. Quantitatively, model 1) yields $\Delta\chi^2 = 98$, \(\Delta\mathrm{BIC} = 92\) (Object~C) and $\Delta\chi^2 = 16$, \(\Delta\mathrm{BIC} = 10\) (Object~F), model 2) results in $\Delta\chi^2 = 107$, \(\Delta\mathrm{BIC} = 89\) (Object~C) and $\Delta\chi^2 = 66 $, \(\Delta\mathrm{BIC}= 54\) (Object~F), while for a three-Gaussian model (model 3) we report $\Delta\chi^2 = 30$, \(\Delta\mathrm{BIC} =  34\) and \(\Delta\mathrm{BIC} = 39\), $\Delta$BIC = 37 for Objects~C and F respectively. These values indicate that the fiducial model provides a statistically better description of the data than its alternatives, and hence that the presence of a narrow component and the measured narrow fluxes are not an obvious artifact of the chosen broad-line parameterisation.

Finally, it could be argued that the deviations from Case B could also be alleviated by splitting the narrow emission into two separate components and assuming a different $A_V$ for each. This, however, cannot solve the problem here, as the flat ratios point to missing \ha flux in the case of Object~C, or a combination of suppressed \ha and enhanced higher order Balmer emission for Object~F. These scenarios cannot be reconciled by combining two Case B components with different extinctions. This is because dust extinction affects higher order Balmer lines more strongly than \ha; any mixture of Case~B spectra subject to different $A_V$ values, will always yield observed \ha/\hb ratios greater than, or equal to the intrinsic Case~B value.

\subsubsection{The case for a hidden absorption line}
\begin{figure}[h!]
    \centering
    \includegraphics[width=\hsize]{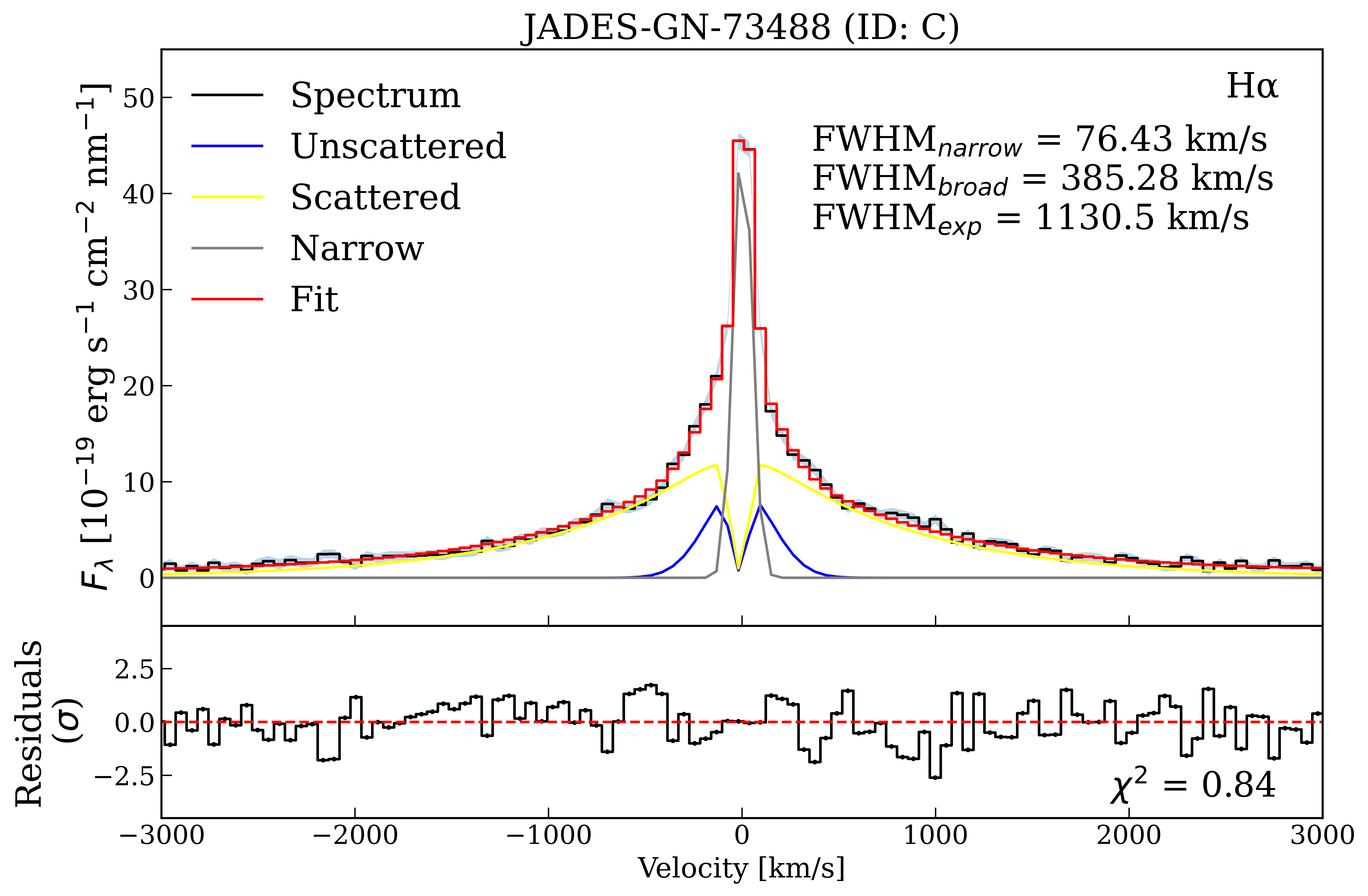}
       \caption{The \ha fit including unresolved absorption for JADES-GN-73488 (Object~C).}
          \label{fig-ha-c-abs}
\end{figure}

\begin{figure}[h!]
    \centering
    \includegraphics[width=\hsize]{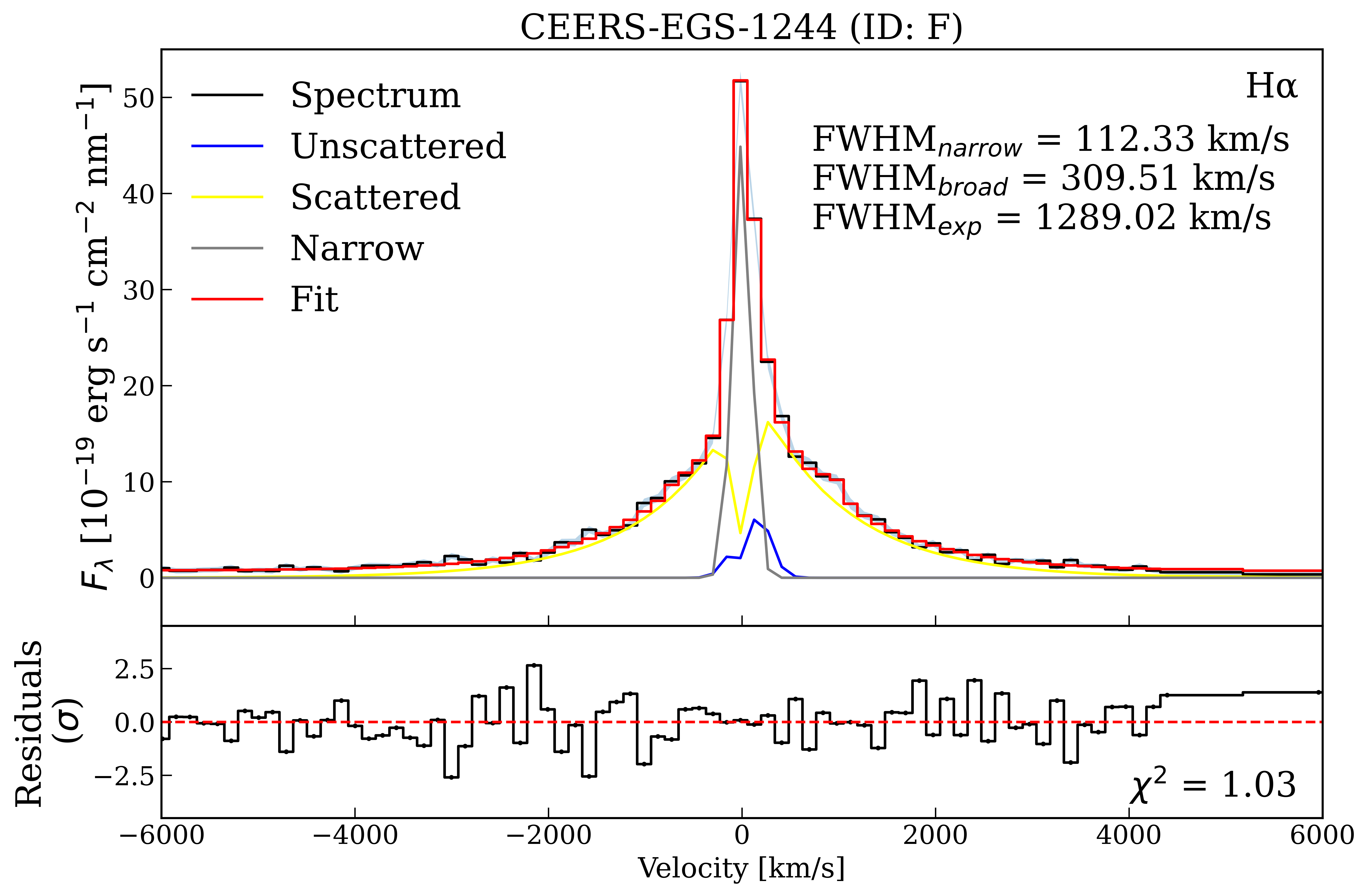}
       \caption{The \ha fit including unresolved absorption for CEERS-EGS-1244 (Object~F).}
          \label{fig-ha-f-abs}
\end{figure}

In principle, the flat narrow-line Balmer ratios could be reconciled if the flux in the narrow \ha line is underestimated. \citet{DEugenio2025} use multi-epoch \jwst NIRSpec/MSA and NIRSpec/IFU spectra of an LRD at $z \simeq 5.077$, observed by the JADES \citep{Eisenstein2023} and BlackTHUNDER (PID~5015;
PIs H.~Übler and R.~Maiolino) surveys. Among other results, they note that the G395H high-resolution data show absorption features almost at rest ($v_{abs} = -13$\,km\,s\(^{-1}\)), which is not detected in the G395M medium resolution data of the same epoch. If such a feature was included in our model, it would reduce the inferred broad flux of the examined emission lines, while increasing the narrow line flux by the same amount. Motivated by this argument, as well as private communication with Jorryt Matthee confirming that unresolved absorption emerges occasionally when re-observing LRDs with higher resolution or longer exposure time, we refit the \ha lines of objects~C and F with an extra component corresponding to undetected/unresolved absorption affecting their broad emission component. The amplitude of an unresolved absorber and that of the narrow Gaussian are of course degenerate in such an explanation. Therefore, we adopted a conservative prior that forces the narrow amplitude to be at least as large as in the fiducial (no-absorption) fit; this choice biases the solution towards transferring flux from the broad to the narrow component. The fit results are shown in Figures~\ref{fig-ha-c-abs} and \ref{fig-ha-f-abs}. 
Upon applying absorption, these objects become consistent with Case~B. Specifically, object~C yields \ha/\hb $= 2.97^{+0.99}_{-0.03}$, while for Object~F we have \ha/\hb $= 3.11^{+0.80}_{-0.85}$. Repeating the methodology of section \ref{sec-res-extinctioncurves} yields a narrow $A_V = 0.12^{+0.017}_{-0.18}$ and $A_V = 0.25^{+0.18}_{-0.17}$ for Objects C and F respectively, both consistent with no extinction. By contrast, the broad $A_V$ values are consistent within $1\sigma$ of the fiducial model values for both objects, remaining consistent with Case B and dust extinction. 
The fits with and without an absorption component are statistically indistinguishable ($\chi^2_{\rm{fiducial}} - \chi^2_{\rm{absorption}} <3$); however, the absorption model introduces one additional free parameter (the absorption amplitude) and is obviously penalised by the BIC, which increases by 2 and 6 for Objects~C and~F, respectively. We nevertheless caution that this inclusion is artificial, in the sense that it requires constraining the priors of the fit so that the flux in the narrow component is increased. When allowing the parameters to be completely free, the best result corresponds to the fiducial model. In addition, the data used in the analysis of JADES-GN-73488 (Object~C) are high resolution, with the object being the second highest signal-to-noise ratio object in our sample \citep{Rusakov2025}. In that sense, at least for Object~C, such a hidden feature seems unlikely. 

\subsubsection{Comparison with other observations}
Recent work probing the new population of high redshift objects revealed by \jwst has revealed \ha/\hb ratios lower than 2.7. For example, \citet{Zhang2025-NarrowLRDS} explore the Balmer decrement of the narrow component of five narrow-line LRDs and find two objects whose decrements are flat enough to imply $A_V \sim -0.5$. Nevertheless, these decrements are not as flat as those presented in this work. \citet{McClymont2025} and  \citet{Sun2025-flatdecs} examine a sample of star-forming, narrow-line \jwst-detected galaxies, finding lower \ha/\hb ratios than predicted by Case~B: several objects detected both in lower resolution PRISM and medium resolution grating spectra have \ha/\hb $< 2$, consistent with our results. 

In summary, we have tested whether the unusually flat narrow Balmer decrements in Objects~C and F can be attributed to modeling biases. Our model comparisons favour the fiducial model over the alternatives explored. Given that similarly low narrow \ha/\hb ratios have been reported in other LRDs and in high-redshift star-forming systems, we consider the departure from Case~B in these narrow components to be a possible solution, although surprising. In Section~\ref{sec-agn-vs-sf} we discuss possible physical explanations and the wider implications of this result.

\subsection{Deviation from Case~B in the broad component of RUBIES-EGS-49140}
\label{sec-objectE}
The broad component of RUBIES-EGS-49140 (Object~E) exhibits a clear deviation from Case~B recombination; its \hg/\ha ratio deviates from Case~B by $\sim 2\sigma$ while the \hd/\ha ratio is more than $5\sigma$ lower than the value expected for Case~B and dust extinction. 

As shown in Figure~\ref{fig-ratios-main}, the \hg/\ha and \hd/\ha ratios are lower than what the \ha/\hb ratio would imply under Case~B conditions with extinction. The fact that RUBIES-EGS-49140 deviates from Case~B has been recently reported also in \citet{D'Eugenio2025deviation}, using the broad \ha, \hb and \hg line ratios. Before concluding in a deviation from Case~B recombination, we also tested whether a steeper extinction curve would alleviate the tensions. Specifically, a \citet{Fitzpatrick1999} Milky Way extinction curve with a variable $R_V$ was used. We found that not even an $R_V$ of 2 was sufficient to accommodate the rapidly decreasing flux in the broad component of the Balmer lines, leading to the conclusion that an unphysically steep curve is needed to explain the anomalous Balmer emission of RUBIES-EGS-49140.

This object shows a strong absorption feature in the cores of both the \ha and \hb, in a form that does not resemble a P~Cygni profile (see Figure~\ref{fig:objectE}). The broad component of \hg is too faint and contaminated by the [O\,\textsc{iii}]\,4363\AA\ line for the absorption profile to be detected, while the \hd line is entirely undetected. We model the \ha profile with a Gaussian absorber of FWHM $\approx$ 350\,km\,s\(^{-1}\), redshifted by 47\,km\,s\(^{-1}\) with respect to its broad component, while \hb requires a Gaussian absorber of FWHM $\approx$ 520\,km\,s\(^{-1}\), \emph{blueshifted} by \(-79\)\,km\,s\(^{-1}\). Each absorber removes a significant fraction of the broad flux; 10\% and 21\% for \ha and \hb respectively, with the absorption depth increasing with decreasing wavelength, as also found in \cite{D'Eugenio2025deviation}. The small velocity offsets of these absorbers strongly suggest that the absorbing gas is also the emitting gas, pointing to a self-absorption scenario \citep{Rusakov2025}, rather than an unrelated fast outflow or a foreground cloud \citep[as examined by][]{juodzbalis2024}. The latter scenario can also be dismissed from the fact that the absorption profiles of these lines are different from one another. 

These self-absorption features imply that the Balmer lines are optically thick, violating a core assumption of Case~B. In the optically thick limit the Balmer lines become depth-dependent: due to a varying optical depth per line, each transition probes a different escape depth in the emitting cloud, so \ha and \hb are naturally expected to have systematically different absorption profiles. Recent observational studies argue that LRDs are embedded in gas cocoons exhibiting high volume and column densities \citep{Rusakov2025,Naidu2025,deGraaff2025Cliff,Inayoshi2025, Taylor2025}, high enough to give rise to non-Case~B Balmer emission \citep{Netzer1975, Drake1980}. We then conclude that the most plausible scenario for the absorption features in RUBIES-EGS-49140, is self–absorption in an optically thick broad line cocoon. In the densities and temperatures required for the Balmer lines to be optically thick, collisional excitation to $n>2$ also affects the hydrogen level populations; quantifying the intrinsic Balmer ratios in this object requires full radiative transfer calculations. 

\subsection{Reddened AGN with a star-forming host galaxy?}
\label{sec-agn-vs-sf}
Recent \jwst studies of high-$z$ AGN and LRDs reveal  dust obscuration, while the host galaxies, when observable, often show little-to-no reddening in the UV and the narrow components of recombination lines. For example, \cite{Killi2024, Rinaldi2024} analyzed J0657\_1045 at $z \approx 4.53$ and found that it was unresolved in red/NIR filters but extended in blue/UV filters. Their spectral fits required two separate continua; a heavily reddened power-law ($A_V \approx 5.6$) dominating the optical spectrum, with a nearly dust-free power-law ($A_V \approx 0.38$) describing the UV. The above results were confirmed by their emission line ratio fits. Across LRD samples, the narrow-line regions often show unattenuated Balmer decrements, contrary to the broad lines \citep{ Matthee2024, Brooks2025, Zhang2025-NarrowLRDS}. Interestingly enough, in some cases, flat narrow \ha/\hb -- some as low as 2 -- are reported in the literature for LRDs \citep{Brooks2025}. 

A number of theoretical work reinforces the above two-component picture. \citet{Volonteri2025} use the \texttt{OBELISK} \citep{Trebitsch2025} cosmological simulation to identify simulated sources with colors similar to those observed in LRDs. They find that such color-selected systems are better described by a dusty AGN dominating the red NIRCam filters, while the blue NIRCam filters are dominated by emission from the host galaxy. In fact, they argue that a galaxy without an active nucleus would not be able to satisfy the color selection criteria of LRDs. Finally, any composite object consisting of an AGN and a star-forming host that has the properties of LRDs, tends to show on average $A_V$~=~1--5, necessary to produce the red optical slopes. This range is in broad agreement with our findings. Similarly, \cite{LaChance2025} use the \texttt{Astrid} \citep{Ni2022, Bird2022, Ni2024} cosmological hydrodynamic simulation to hunt for LRDs. They conclude that the components that make up LRDs include 1) a recently quenched star-forming host galaxy, 2) a luminous AGN accreting at a high Eddington ratio and 3) adequate dust attenuation so that the red optical slope is dominated by the AGN while the UV is dominated by stars. 

In this work, for the first time, we include the analysis of Balmer lines beyond \ha and \hb, yielding more accurate estimates of $A_V$ and allowing us to discriminate sources that do not follow Case~B recombination. Assuming the validity of Case~B, for objects whose \hg and \hd lines are upper limits, we find that almost all objects show a significant difference between the extinction derived from the narrow lines and that required by the broad lines, with an average $A_{V,\mathrm{narrow}} \approx 0$ and $A_{V,\mathrm{broad}} \approx 4.13$. Quantitatively, using an SMC dust law for which the UV extinction, \(A_{1500\,\mathrm{\AA}} \approx 5.4\times A_V\) \citep{Nakazato2024}, the average broad $A_V$ in our sample implies a UV transmitted fraction of $f_\mathrm{UV} = 10^{-0.4\,A_\mathrm{UV}} \approx 10^{-2.2\,A_V} \approx 10^{-9}$. The above fraction would suggest that no UV light escapes unless the dust law in the UV is intrinsically very flat. Clumpiness does not really work, as it would produce a flat curve in the optical as well, which is contrary to the reddening we are measuring. In a dust-dominated scenario, our results favour the notion that the broad/optical-NIR colors arise from a central engine heavily extinguished by dust, with the blue/UV colors emerging from a spatially distinct region with minimal dust attenuation.

\subsection{Lack of dust emission in the Infrared and the implications of departures from Case~B}
\label{sec-ir-problem}
However, while we find in this work that a high percentage of our objects, yield broad line Balmer ratios that show consistency with high dust extinction, an enduring puzzle in the literature is the mismatch between the Balmer decrement-inferred obscuration in LRDs and their IR emission. \jwst studies with MIRI reveal that LRDs are too faint in the rest-frame NIR, showing no traces of a compact, warm dusty torus emission, commonly found in AGN \citep{Akins2024, PerezGonzalez2024,Leung2024,Setton2025}. Moreover, invoking a dusty torus model to fit the optical/NIR spectra of LRDs yields a dust temperature of $T_b \sim 2500$\,K \citep{Killi2024}, which not only exceeds the sublimation temperatures of carbon-based dust ($T_b \sim 2000$\,K, \citealt{Kobayashi2009}), but is also significantly higher than typical dust torus inner temperatures of Type~1 AGN ($T_b \sim 1400$\,K, \citealt{Kishimoto2007, Honig2010}). In order to reconcile the constraints set by MIRI on hot dust, the dust needs to be spatially extended \citep{Li2025}, or distributed in clumps \citep{Honig2010, Casey2024}, shifting the peak of the black-body dust distribution to the mid-infrared.

Recent work has established that LRDs are embedded in extremely dense gas, with column densities of the order of $N_H \sim 10^{24} - 10^{26},\rm{cm^{-2}}$ \citep{Rusakov2025, Kido2025, Ji2025, Naidu2025}. Under such conditions, \ha emission can be significantly enhanced by self-absorption or by collisional excitation to the $n = 3$ level, producing artificially steep Balmer decrements. Interpreting these steep ratios under the assumption of pure Case~B recombination leads to an overestimation of the dust extinction in their spectra. If these high-density effects are indeed at play in LRDs, then part of the tension between the large inferred $A_V$ values and the relatively weak IR emission could be alleviated. Furthermore, the redness of their continuum slopes are then at least partly explained by the nebular continua. A comprehensive analysis of LRDs will therefore require deeper spectroscopic observations -- sensitive enough to detect the broad components of higher-order Balmer lines (\hg and above) -- combined with modelling that explicitly accounts for high-density effects. Even without detections of \hd and higher-order lines, objects whose spectra are strongly shaped by high-density gas should still exhibit clear departures from Case~B, as seen in RUBIES-EGS-49140 (Object~E). Several more cases as clear or clearer than Object~E in the data would militate against extinction as the primary cause of large Balmer decrements.

\subsubsection{Negative Balmer decrements}
Balmer decrements lower than the Case~B value are observed in galaxies at low~$z$ (e.g.\ \citealt{Atek2009, Scarlata2024}) and high~$z$ (e.g.\ \citealt{Cameron2024, Topping2024,McClymont2025, Sun2025-flatdecs}), and could be associated with density-bounded nebulae in starbursts \citep{McClymont2025}. In density bounded regions, the assumption that Lyman photons are instantaneously absorbed and thus pump the corresponding Balmer lines, breaks down. As a result, the emerging total Balmer emission looks anomalous because the various line ratios change as a function of cloud depth, only reaching Case~B when the Lyman lines become optically thick \citep{McClymont2025}. 
The objects in our sample with narrow component Balmer decrements \ha/\hb $\approx 2$ (Objects~C and F) are also the bluest objects in the sample \citep[see Extended Data Figure ~2 in][]{Rusakov2025}. If these blue colors are indeed not powered by the black hole itself, as hinted by the results in this work, then their coincidence with flat Balmer ratios -- observed in star forming galaxies at high-$z$ -- points toward a star-forming origin for the blue component of LRDs. 

Alternatively, when taking into account the scattering of Balmer photons due to increased optical depth, Balmer self-absorption and collisional de-excitation, one can reach \ha/\hb as flat as the values found in this work \citep[see][and references therein]{Netzer1975, Drake1980,  Scarlata2024}. However, the processes that arise from increased Balmer optical depth are important at very high densities (log$n_e > 10$, \citealt{Netzer1975, Drake1980}) and overall physical conditions which do not resemble a typical extended NLR. We therefore argue that a density bounded nebula origin for these flat narrow decrements or unresolved absorption in the broad lines are more plausible scenarios. The interested reader is referred to  \cite{Netzer1975,Drake1980} and \cite{Scarlata2024} for a very thorough discussion on the validity of Case~B recombination and the conditions which lead to its breakdown.

High-$z$ AGN and LRDs seem to be a remarkably heterogeneous population \citep{Rinaldi2024, Carranza2025}. We expect that a combination of physically distinct regions (star-forming and AGN-like) may need to be invoked to explain their UV spectra, with the relative contribution of each component varying from object to object.

\section{Summary}
\label{sec-summary}
In this work, we analyzed the Balmer line ratios (\ha, \hb, \hg, \hd) of a sample of high-$z$ AGN compiled in \cite{Rusakov2025}, decomposing each line into their respective narrow and broad components. By incorporating higher-order Balmer lines, we tested for departures from Case~B recombination and quantified dust extinction with improved accuracy. Our results show that Case~B cannot be robustly ruled out for all examined narrow components, and 6 out of 7 broad components, given the non-detections of \hg and \hd. To be consistent with Case~B, we find that the broad components would require heavy obscuration ($A_V \sim 4.13$) while the narrow components are on average dust-free ($A_V \sim 0$). Regardless of the origin of the difference, this dichotomy strongly supports a two-component scenario, where the broad lines (and possibly the optical/NIR emission) originate from a high column density AGN, while the narrow lines (and possibly the UV continuum) arise from a low-dust narrow line region or star-formation. 

Our results also show clear evidence of deviations from Case~B recombination; out of only two objects with a detected broad component in \hg, one (RUBIES-EGS-49140) shows a significant departure from Case~B, with the broad \hd/\ha ratio lying more than $5\sigma$ below the Case~B prediction. Furthermore, we present two objects (JADES-GN-73488 and CEERS-EGS-1244) whose narrow \ha/\hb are surprisingly flat ($\sim 2$), indicative of emission arising from density bounded nebulae in starbursts or unresolved absorption in their broad lines. We point out that it is also possible that other objects in our sample intrinsically deviate from Case~B, but such deviations remain hidden because their \hg and \hd broad components are only constrained by upper limits. To determine whether such deviations are common among high-redshift AGN, future observations must achieve higher signal-to-noise and spectral resolution, enabling robust detection of the broad components of higher-order Balmer lines.

Finally, if the observed ratios truly reflect Case~B with heavy dust extinction, then intrinsic luminosities -- and thus black hole masses inferred from virial methods -- would likely be underestimated. An $A_V$ of 5, implies a flux correction factor of $10^{0.4A_V} = 100$; since log$M_{BH} \propto 0.5$log$L_{\mathrm{H}\alpha}$ \citep{Greene2005}, in such a case the inferred mass would increase by about 1\,dex due to the dust correction. Conversely, in cases where Case~B does not hold, any estimate based on Balmer emission becomes unreliable, and detailed radiative transfer modeling on an object-by-object basis is required. 

\begin{acknowledgements}
    We would like to thank Jorryt Matthee for inspiring discussions.  
    The Cosmic Dawn Center (DAWN) is funded by the Danish National Research Foundation under grant DNRF140. The data products presented herein were retrieved from the DAWN \jwst\ Archive (DJA). DJA is an initiative of the Cosmic Dawn Center. DW, GPN and AS are co-funded by the European Union (ERC, HEAVYMETAL, 101071865). Views and opinions expressed are, however, those of the authors only and do not necessarily reflect those of the European Union or the European Research Council. Neither the European Union nor the granting authority can be held responsible for them. VR, is funded by the ERC Advanced Investigator Grant EPOCHS (788113). KEH acknowledges funding from the Swiss State Secretariat for Education, Research and Innovation (SERI). This work is based in part on observations made with the NASA/ESA/CSA James Webb Space Telescope. The data were obtained from the Mikulski Archive for Space Telescopes (MAST) at the Space Telescope Science Institute, which is operated by the Association of Universities for Research in Astronomy, Inc., under NASA contract NAS 5-03127 for \jwst.

%
\end{acknowledgements}


\begin{thebibliography}{77}
\expandafter\ifx\csname natexlab\endcsname\relax\def\natexlab#1{#1}\fi

\bibitem[{{Akins} {et~al.}(2024){Akins}, {Casey}, {Lambrides}, {Allen}, {Andika}, {Brinch}, {Champagne}, {Cooper}, {Ding}, {Drakos}, {Faisst}, {Finkelstein}, {Franco}, {Fujimoto}, {Gentile}, {Gillman}, {Gozaliasl}, {Harish}, {Hayward}, {Hirschmann}, {Ilbert}, {Kartaltepe}, {Kocevski}, {Koekemoer}, {Kokorev}, {Liu}, {Long}, {McCracken}, {McKinney}, {Onoue}, {Paquereau}, {Renzini}, {Rhodes}, {Robertson}, {Shuntov}, {Silverman}, {Tanaka}, {Toft}, {Trakhtenbrot}, {Valentino}, \& {Zavala}}]{Akins2024}
{Akins}, H.~B., {Casey}, C.~M., {Lambrides}, E., {et~al.} 2024, arXiv e-prints, arXiv:2406.10341

\bibitem[{{Ananna} {et~al.}(2024){Ananna}, {Bogd{\'a}n}, {Kov{\'a}cs}, {Natarajan}, \& {Hickox}}]{Ananna2024}
{Ananna}, T.~T., {Bogd{\'a}n}, {\'A}., {Kov{\'a}cs}, O.~E., {Natarajan}, P., \& {Hickox}, R.~C. 2024, \apjl, 969, L18

\bibitem[{{Atek} {et~al.}(2009){Atek}, {Kunth}, {Schaerer}, {Hayes}, {Deharveng}, {{\"O}stlin}, \& {Mas-Hesse}}]{Atek2009}
{Atek}, H., {Kunth}, D., {Schaerer}, D., {et~al.} 2009, \aap, 506, L1

\bibitem[{{Bird} {et~al.}(2022){Bird}, {Ni}, {Di Matteo}, {Croft}, {Feng}, \& {Chen}}]{Bird2022}
{Bird}, S., {Ni}, Y., {Di Matteo}, T., {et~al.} 2022, \mnras, 512, 3703

\bibitem[{{Brooks} {et~al.}(2025){Brooks}, {Simons}, {Trump}, {Taylor}, {Bagley}, {Backhaus}, {Davis}, {Buat}, {Cleri}, {de la Vega}, {Finkelstein}, {Hirschmann}, {Holwerda}, {Kocevski}, {Koekemoer}, {Lucas}, {Pacucci}, \& {Seill{\'e}}}]{Brooks2025}
{Brooks}, M., {Simons}, R.~C., {Trump}, J.~R., {et~al.} 2025, \apj, 986, 177

\bibitem[{{Cameron} {et~al.}(2024){Cameron}, {Katz}, {Witten}, {Saxena}, {Laporte}, \& {Bunker}}]{Cameron2024}
{Cameron}, A.~J., {Katz}, H., {Witten}, C., {et~al.} 2024, \mnras, 534, 523

\bibitem[{{Carranza-Escudero} {et~al.}(2025){Carranza-Escudero}, {Conselice}, {Adams}, {Harvey}, {Austin}, {Behroozi}, {Ferreira}, {Ormerod}, {Duan}, {Trussler}, {Li}, {Westcott}, {Windhorst}, {Coe}, {Cohen}, {Cheng}, {Driver}, {Frye}, {Furtak}, {Grogin}, {Hathi}, {Jansen}, {Koekemoer}, {Marshall}, {O'Brien}, {Pirzkal}, {Polletta}, {Robotham}, {Rutkowski}, {Summers}, {Wilkins}, {Willmer}, {Yan}, \& {Zitrin}}]{Carranza2025}
{Carranza-Escudero}, M., {Conselice}, C.~J., {Adams}, N., {et~al.} 2025, arXiv e-prints, arXiv:2506.04004

\bibitem[{Casey {et~al.}(2024)Casey, Akins, Kokorev, McKinney, Cooper, Long, Franco, \& Manning}]{Casey2024}
Casey, C.~M., Akins, H.~B., Kokorev, V., {et~al.} 2024, The Astrophysical Journal Letters, 975, L4

\bibitem[{{Chang} {et~al.}(2025){Chang}, {Gronke}, {Matthee}, \& {Mason}}]{Chang2025}
{Chang}, S.-J., {Gronke}, M., {Matthee}, J., \& {Mason}, C. 2025, arXiv e-prints, arXiv:2508.08768

\bibitem[{{de Graaff} {et~al.}(2025{\natexlab{a}}){de Graaff}, {Brammer}, {Weibel}, {Lewis}, {Maseda}, {Oesch}, {Bezanson}, {Boogaard}, {Cleri}, {Cooper}, {Gottumukkala}, {Greene}, {Hirschmann}, {Hviding}, {Katz}, {Labb{\'e}}, {Leja}, {Matthee}, {McConachie}, {Miller}, {Naidu}, {Price}, {Rix}, {Setton}, {Suess}, {Wang}, {Whitaker}, \& {Williams}}]{deGraaff2025RUBIES}
{de Graaff}, A., {Brammer}, G., {Weibel}, A., {et~al.} 2025{\natexlab{a}}, \aap, 697, A189

\bibitem[{{de Graaff} {et~al.}(2024){de Graaff}, {Rix}, {Carniani}, {Suess}, {Charlot}, {Curtis-Lake}, {Arribas}, {Baker}, {Boyett}, {Bunker}, {Cameron}, {Chevallard}, {Curti}, {Eisenstein}, {Franx}, {Hainline}, {Hausen}, {Ji}, {Johnson}, {Jones}, {Maiolino}, {Maseda}, {Nelson}, {Parlanti}, {Rawle}, {Robertson}, {Tacchella}, {{\"U}bler}, {Williams}, {Willmer}, \& {Willott}}]{deGraaff2024}
{de Graaff}, A., {Rix}, H.-W., {Carniani}, S., {et~al.} 2024, \aap, 684, A87

\bibitem[{{de Graaff} {et~al.}(2025{\natexlab{b}}){de Graaff}, {Rix}, {Naidu}, {Labbe}, {Wang}, {Leja}, {Matthee}, {Katz}, {Greene}, {Hviding}, {Baggen}, {Bezanson}, {Boogaard}, {Brammer}, {Dayal}, {van Dokkum}, {Goulding}, {Hirschmann}, {Maseda}, {McConachie}, {Miller}, {Nelson}, {Oesch}, {Setton}, {Shivaei}, {Weibel}, {Whitaker}, \& {Williams}}]{deGraaff2025Cliff}
{de Graaff}, A., {Rix}, H.-W., {Naidu}, R.~P., {et~al.} 2025{\natexlab{b}}, arXiv e-prints, arXiv:2503.16600

\bibitem[{{D'Eugenio} {et~al.}(2025{\natexlab{a}}){D'Eugenio}, {Juod{\v{z}}balis}, {Ji}, {Scholtz}, {Maiolino}, {Carniani}, {Perna}, {Mazzolari}, {{\"U}bler}, {Arribas}, {Bhatawdekar}, {Bunker}, {Cresci}, {Curtis-Lake}, {Hainline}, {Inayoshi}, {Isobe}, {Johnson}, {Jones}, {Looser}, {Nelson}, {Parlanti}, {Pusk{\'a}s}, {Rinaldi}, {Robertson}, {Rodr{\'\i}guez Del Pino}, {Shivaei}, {Sun}, {Tacchella}, {Venturi}, {Volonteri}, {Williams}, {Willmer}, {Willott}, \& {Witstok}}]{DEugenio2025}
{D'Eugenio}, F., {Juod{\v{z}}balis}, I., {Ji}, X., {et~al.} 2025{\natexlab{a}}, arXiv e-prints, arXiv:2506.14870

\bibitem[{{D'Eugenio} {et~al.}(2025{\natexlab{b}}){D'Eugenio}, {Nelson}, {Ji}, {Baggen}, {Greene}, {Labb{\'e}}, {Pezzulli}, {Brown}, {Maiolino}, {Matthee}, {Terlevich}, {Terlevich}, {Torralba}, \& {Carniani}}]{D'Eugenio2025deviation}
{D'Eugenio}, F., {Nelson}, E., {Ji}, X., {et~al.} 2025{\natexlab{b}}, arXiv e-prints, arXiv:2510.00101

\bibitem[{{Drake} \& {Ulrich}(1980)}]{Drake1980}
{Drake}, S.~A. \& {Ulrich}, R.~K. 1980, \apjs, 42, 351

\bibitem[{{Eisenstein} {et~al.}(2023){Eisenstein}, {Willott}, {Alberts}, {Arribas}, {Bonaventura}, {Bunker}, {Cameron}, {Carniani}, {Charlot}, {Curtis-Lake}, {D'Eugenio}, {Endsley}, {Ferruit}, {Giardino}, {Hainline}, {Hausen}, {Jakobsen}, {Johnson}, {Maiolino}, {Rieke}, {Rieke}, {Rix}, {Robertson}, {Stark}, {Tacchella}, {Williams}, {Willmer}, {Baker}, {Baum}, {Bhatawdekar}, {Boyett}, {Chen}, {Chevallard}, {Circosta}, {Curti}, {Danhaive}, {DeCoursey}, {de Graaff}, {Dressler}, {Egami}, {Helton}, {Hviding}, {Ji}, {Jones}, {Kumari}, {L{\"u}tzgendorf}, {Laseter}, {Looser}, {Lyu}, {Maseda}, {Nelson}, {Parlanti}, {Perna}, {Pusk{\'a}s}, {Rawle}, {Rodr{\'\i}guez Del Pino}, {Sandles}, {Saxena}, {Scholtz}, {Sharpe}, {Shivaei}, {Silcock}, {Simmonds}, {Skarbinski}, {Smit}, {Stone}, {Suess}, {Sun}, {Tang}, {Topping}, {{\"U}bler}, {Villanueva}, {Wallace}, {Whitler}, {Witstok}, \& {Woodrum}}]{Eisenstein2023}
{Eisenstein}, D.~J., {Willott}, C., {Alberts}, S., {et~al.} 2023, arXiv e-prints, arXiv:2306.02465

\bibitem[{{Feroz} {et~al.}(2009){Feroz}, {Hobson}, \& {Bridges}}]{Feroz2009}
{Feroz}, F., {Hobson}, M.~P., \& {Bridges}, M. 2009, \mnras, 398, 1601

\bibitem[{Finkelstein {et~al.}(2023)Finkelstein, Bagley, Ferguson, Wilkins, Kartaltepe, Papovich, Yung, Arrabal~Haro, Behroozi, Dickinson, Kocevski, Koekemoer, Larson, Le~Bail, Morales, Pérez-González, Burgarella, Davé, Hirschmann, Somerville, Wuyts, Bromm, Casey, Fontana, Fujimoto, Gardner, Giavalisco, Grazian, Grogin, Hathi, Hutchison, Jha, Jogee, Kewley, Kirkpatrick, Long, Lotz, Pentericci, Pierel, Pirzkal, Ravindranath, Ryan, Trump, Yang, Bhatawdekar, Bisigello, Buat, Calabrò, Castellano, Cleri, Cooper, Croton, Daddi, Dekel, Elbaz, Franco, Gawiser, Holwerda, Huertas-Company, Jaskot, Leung, Lucas, Mobasher, Pandya, Tacchella, Weiner, \& Zavala}]{Finkelstein2023}
Finkelstein, S.~L., Bagley, M.~B., Ferguson, H.~C., {et~al.} 2023, The Astrophysical Journal Letters, 946, L13

\bibitem[{{Fitzpatrick}(1999)}]{Fitzpatrick1999}
{Fitzpatrick}, E.~L. 1999, \pasp, 111, 63

\bibitem[{Foreman-Mackey(2016)}]{corner}
Foreman-Mackey, D. 2016, The Journal of Open Source Software, 1, 24

\bibitem[{{Gloudemans} {et~al.}(2025){Gloudemans}, {Duncan}, {Eilers}, {Farina}, {Harikane}, {Inayoshi}, {Lambrides}, \& {Vardoulaki}}]{Gloudemans2025}
{Gloudemans}, A.~J., {Duncan}, K.~J., {Eilers}, A.-C., {et~al.} 2025, \apj, 986, 130

\bibitem[{Gordon(2024{\natexlab{a}})}]{lmc}
Gordon, K.~D. 2024{\natexlab{a}}, Journal of Open Source Software, 9, 7023

\bibitem[{Gordon(2024{\natexlab{b}})}]{Gordon2024-dust_ext}
Gordon, K.~D. 2024{\natexlab{b}}, Journal of Open Source Software, 9, 7023

\bibitem[{{Gordon} {et~al.}(2023){Gordon}, {Clayton}, {Decleir}, {Fitzpatrick}, {Massa}, {Misselt}, \& {Tollerud}}]{mw}
{Gordon}, K.~D., {Clayton}, G.~C., {Decleir}, M., {et~al.} 2023, \apj, 950, 86

\bibitem[{{Gordon} {et~al.}(2024){Gordon}, {Fitzpatrick}, {Massa}, {Bohlin}, {Chastenet}, {Murray}, {Clayton}, {Lennon}, {Misselt}, \& {Sandstrom}}]{Gordon2024-SMC}
{Gordon}, K.~D., {Fitzpatrick}, E.~L., {Massa}, D., {et~al.} 2024, \apj, 970, 51

\bibitem[{{Greene} \& {Ho}(2005)}]{Greene2005}
{Greene}, J.~E. \& {Ho}, L.~C. 2005, \apj, 630, 122

\bibitem[{{Greene} {et~al.}(2024){Greene}, {Labbe}, {Goulding}, {Furtak}, {Chemerynska}, {Kokorev}, {Dayal}, {Volonteri}, {Williams}, {Wang}, {Setton}, {Burgasser}, {Bezanson}, {Atek}, {Brammer}, {Cutler}, {Feldmann}, {Fujimoto}, {Glazebrook}, {de Graaff}, {Khullar}, {Leja}, {Marchesini}, {Maseda}, {Matthee}, {Miller}, {Naidu}, {Nanayakkara}, {Oesch}, {Pan}, {Papovich}, {Price}, {van Dokkum}, {Weaver}, {Whitaker}, \& {Zitrin}}]{Greene2024}
{Greene}, J.~E., {Labbe}, I., {Goulding}, A.~D., {et~al.} 2024, \apj, 964, 39

\bibitem[{{Harikane} {et~al.}(2023){Harikane}, {Zhang}, {Nakajima}, {Ouchi}, {Isobe}, {Ono}, {Hatano}, {Xu}, \& {Umeda}}]{Harikane2023}
{Harikane}, Y., {Zhang}, Y., {Nakajima}, K., {et~al.} 2023, \apj, 959, 39

\bibitem[{{Heintz} {et~al.}(2025){Heintz}, {Brammer}, {Watson}, {Oesch}, {Keating}, {Hayes}, {Abdurro'uf}, {Arellano-C{\'o}rdova}, {Carnall}, {Christiansen}, {Cullen}, {Dav{\'e}}, {Dayal}, {Ferrara}, {Finlator}, {Fynbo}, {Flury}, {Gelli}, {Gillman}, {Gottumukkala}, {Gould}, {Greve}, {Hardin}, {Hsiao}, {Hutter}, {Jakobsson}, {Killi}, {Khosravaninezhad}, {Laursen}, {Lee}, {Magdis}, {Matthee}, {Naidu}, {Narayanan}, {Pollock}, {Prescott}, {Rusakov}, {Shuntov}, {Sneppen}, {Smit}, {Tanvir}, {Terp}, {Toft}, {Valentino}, {Vijayan}, {Weaver}, {Wise}, \& {Witstok}}]{Heintz2025}
{Heintz}, K.~E., {Brammer}, G.~B., {Watson}, D., {et~al.} 2025, \aap, 693, A60

\bibitem[{{H{\"o}nig} \& {Kishimoto}(2010)}]{Honig2010}
{H{\"o}nig}, S.~F. \& {Kishimoto}, M. 2010, \aap, 523, A27

\bibitem[{{Inayoshi} \& {Maiolino}(2025)}]{Inayoshi2025}
{Inayoshi}, K. \& {Maiolino}, R. 2025, \apjl, 980, L27

\bibitem[{{Jakobsen} {et~al.}(2022){Jakobsen}, {Ferruit}, {Alves de Oliveira}, {Arribas}, {Bagnasco}, {Barho}, {Beck}, {Birkmann}, {B{\"o}ker}, {Bunker}, {Charlot}, {de Jong}, {de Marchi}, {Ehrenwinkler}, {Falcolini}, {Fels}, {Franx}, {Franz}, {Funke}, {Giardino}, {Gnata}, {Holota}, {Honnen}, {Jensen}, {Jentsch}, {Johnson}, {Jollet}, {Karl}, {Kling}, {K{\"o}hler}, {Kolm}, {Kumari}, {Lander}, {Lemke}, {L{\'o}pez-Caniego}, {L{\"u}tzgendorf}, {Maiolino}, {Manjavacas}, {Marston}, {Maschmann}, {Maurer}, {Messerschmidt}, {Moseley}, {Mosner}, {Mott}, {Muzerolle}, {Pirzkal}, {Pittet}, {Plitzke}, {Posselt}, {Rapp}, {Rauscher}, {Rawle}, {Rix}, {R{\"o}del}, {Rumler}, {Sabbi}, {Salvignol}, {Schmid}, {Sirianni}, {Smith}, {Strada}, {te Plate}, {Valenti}, {Wettemann}, {Wiehe}, {Wiesmayer}, {Willott}, {Wright}, {Zeidler}, \& {Zincke}}]{Jakobsen2022}
{Jakobsen}, P., {Ferruit}, P., {Alves de Oliveira}, C., {et~al.} 2022, \aap, 661, A80

\bibitem[{{Ji} {et~al.}(2025){Ji}, {Maiolino}, {{\"U}bler}, {Scholtz}, {D'Eugenio}, {Sun}, {Perna}, {Turner}, {Arribas}, {Bennett}, {Bunker}, {Carniani}, {Charlot}, {Cresci}, {Curti}, {Egami}, {Fabian}, {Inayoshi}, {Isobe}, {Jones}, {Juod{\v{z}}balis}, {Kumari}, {Lyu}, {Mazzolari}, {Parlanti}, {Robertson}, {Rodr{\'\i}guez Del Pino}, {Schneider}, {Sijacki}, {Tacchella}, {Trinca}, {Valiante}, {Venturi}, {Volonteri}, {Willott}, {Witten}, \& {Witstok}}]{Ji2025}
{Ji}, X., {Maiolino}, R., {{\"U}bler}, H., {et~al.} 2025, arXiv e-prints, arXiv:2501.13082

\bibitem[{Juodžbalis {et~al.}(2024)Juodžbalis, Ji, Maiolino, D'Eugenio, Scholtz, Risaliti, Fabian, Mazzolari, Gilli, Prandoni, Arribas, Bunker, Carniani, Charlot, Curtis-Lake, de~Graaff, Hainline, Parlanti, Perna, Pérez-González, Robertson, Tacchella, Übler, Williams, Willott, \& Witstok}]{juodzbalis2024}
Juodžbalis, I., Ji, X., Maiolino, R., {et~al.} 2024, JADES -- The Rosetta Stone of JWST-discovered AGN: deciphering the intriguing nature of early AGN

\bibitem[{{Kido} {et~al.}(2025){Kido}, {Ioka}, {Hotokezaka}, {Inayoshi}, \& {Irwin}}]{Kido2025}
{Kido}, D., {Ioka}, K., {Hotokezaka}, K., {Inayoshi}, K., \& {Irwin}, C.~M. 2025, arXiv e-prints, arXiv:2505.06965

\bibitem[{{Killi} {et~al.}(2024){Killi}, {Watson}, {Brammer}, {McPartland}, {Antwi-Danso}, {Newshore}, {Coe}, {Allen}, {Fynbo}, {Gould}, {Heintz}, {Rusakov}, \& {Vejlgaard}}]{Killi2024}
{Killi}, M., {Watson}, D., {Brammer}, G., {et~al.} 2024, \aap, 691, A52

\bibitem[{{Kishimoto} {et~al.}(2007){Kishimoto}, {H{\"o}nig}, {Beckert}, \& {Weigelt}}]{Kishimoto2007}
{Kishimoto}, M., {H{\"o}nig}, S.~F., {Beckert}, T., \& {Weigelt}, G. 2007, \aap, 476, 713

\bibitem[{{Kobayashi} {et~al.}(2009){Kobayashi}, {Watanabe}, {Kimura}, \& {Yamamoto}}]{Kobayashi2009}
{Kobayashi}, H., {Watanabe}, S.-i., {Kimura}, H., \& {Yamamoto}, T. 2009, \icarus, 201, 395

\bibitem[{{Kocevski} {et~al.}(2025){Kocevski}, {Finkelstein}, {Barro}, {Taylor}, {Calabr{\`o}}, {Laloux}, {Buchner}, {Trump}, {Leung}, {Yang}, {Dickinson}, {P{\'e}rez-Gonz{\'a}lez}, {Pacucci}, {Inayoshi}, {Somerville}, {McGrath}, {Akins}, {Bagley}, {Bowler}, {Bisigello}, {Carnall}, {Casey}, {Cheng}, {Cleri}, {Costantin}, {Cullen}, {Davis}, {Donnan}, {Dunlop}, {Ellis}, {Ferguson}, {Fujimoto}, {Fontana}, {Giavalisco}, {Grazian}, {Grogin}, {Hathi}, {Hirschmann}, {Huertas-Company}, {Holwerda}, {Illingworth}, {Juneau}, {Kartaltepe}, {Koekemoer}, {Li}, {Lucas}, {Magee}, {Mason}, {McLeod}, {McLure}, {Napolitano}, {Papovich}, {Pirzkal}, {Rodighiero}, {Santini}, {Wilkins}, \& {Yung}}]{Kocevski2025}
{Kocevski}, D.~D., {Finkelstein}, S.~L., {Barro}, G., {et~al.} 2025, \apj, 986, 126

\bibitem[{{Kokubo} \& {Harikane}(2024)}]{Kokubo2024}
{Kokubo}, M. \& {Harikane}, Y. 2024, arXiv e-prints, arXiv:2407.04777

\bibitem[{Koposov {et~al.}(2024)Koposov, Speagle, Barbary, Ashton, Bennett, Buchner, Scheffler, Cook, Talbot, Guillochon, Cubillos, Ramos, Dartiailh, Ilya, Tollerud, Lang, Johnson, jtmendel, Higson, Vandal, Daylan, Angus, patelR, Cargile, Sheehan, Pitkin, Kirk, Leja, joezuntz, \& Goldstein}]{sergey_koposov_2024_12537467}
Koposov, S., Speagle, J., Barbary, K., {et~al.} 2024, joshspeagle/dynesty: v2.1.4

\bibitem[{{LaChance} {et~al.}(2025){LaChance}, {Croft}, {Di Matteo}, {Zhou}, {Pacucci}, {Ni}, {Chen}, \& {Bird}}]{LaChance2025}
{LaChance}, P., {Croft}, R. A.~C., {Di Matteo}, T., {et~al.} 2025, arXiv e-prints, arXiv:2505.20439

\bibitem[{{Leung} {et~al.}(2024){Leung}, {Finkelstein}, {P{\'e}rez-Gonz{\'a}lez}, {Morales}, {Taylor}, {Barro}, {Kocevski}, {Akins}, {Carnall}, {Ch{\'a}vez Ortiz}, {Cleri}, {Cullen}, {Donnan}, {Dunlop}, {Ellis}, {Grogin}, {Hirschmann}, {Koekemoer}, {Kokorev}, {Lucas}, {McLeod}, {Papovich}, \& {Yung}}]{Leung2024}
{Leung}, G. C.~K., {Finkelstein}, S.~L., {P{\'e}rez-Gonz{\'a}lez}, P.~G., {et~al.} 2024, arXiv e-prints, arXiv:2411.12005

\bibitem[{{Li} {et~al.}(2025){Li}, {Inayoshi}, {Chen}, {Ichikawa}, \& {Ho}}]{Li2025}
{Li}, Z., {Inayoshi}, K., {Chen}, K., {Ichikawa}, K., \& {Ho}, L.~C. 2025, \apj, 980, 36

\bibitem[{Maiolino {et~al.}(2025)Maiolino, Risaliti, Signorini, Trefoloni, Juodžbalis, Scholtz, Übler, D’Eugenio, Carniani, Fabian, Ji, Mazzolari, Bertola, Brusa, Bunker, Charlot, Comastri, Cresci, DeCoursey, Egami, Fiore, Gilli, Perna, Tacchella, \& Venturi}]{Maiolino2024Xray}
Maiolino, R., Risaliti, G., Signorini, M., {et~al.} 2025, Monthly Notices of the Royal Astronomical Society, 538, 1921

\bibitem[{{Maiolino} {et~al.}(2024){Maiolino}, {Scholtz}, {Curtis-Lake}, {Carniani}, {Baker}, {de Graaff}, {Tacchella}, {{\"U}bler}, {D'Eugenio}, {Witstok}, {Curti}, {Arribas}, {Bunker}, {Charlot}, {Chevallard}, {Eisenstein}, {Egami}, {Ji}, {Jones}, {Lyu}, {Rawle}, {Robertson}, {Rujopakarn}, {Perna}, {Sun}, {Venturi}, {Williams}, \& {Willott}}]{Maiolino2024}
{Maiolino}, R., {Scholtz}, J., {Curtis-Lake}, E., {et~al.} 2024, \aap, 691, A145

\bibitem[{{Matthee} {et~al.}(2024){Matthee}, {Naidu}, {Brammer}, {Chisholm}, {Eilers}, {Goulding}, {Greene}, {Kashino}, {Labbe}, {Lilly}, {Mackenzie}, {Oesch}, {Weibel}, {Wuyts}, {Xiao}, {Bordoloi}, {Bouwens}, {van Dokkum}, {Illingworth}, {Kramarenko}, {Maseda}, {Mason}, {Meyer}, {Nelson}, {Reddy}, {Shivaei}, {Simcoe}, \& {Yue}}]{Matthee2024}
{Matthee}, J., {Naidu}, R.~P., {Brammer}, G., {et~al.} 2024, \apj, 963, 129

\bibitem[{{Mazzolari} {et~al.}(2024){Mazzolari}, {Gilli}, {Maiolino}, {Prandoni}, {Delvecchio}, {Norman}, {Jimenez-Andrade}, {Belladitta}, {Vito}, {Momjian}, {Chiaberge}, {Trefoloni}, {Signorini}, {Ji}, {D'Amato}, {Risaliti}, {Baldi}, {Fabian}, {{\"U}bler}, {D'Eugenio}, {Scholtz}, {Juod{\v{z}}balis}, {Mignoli}, {Brusa}, {Murphy}, \& {Muxlow}}]{Mazzolari2024}
{Mazzolari}, G., {Gilli}, R., {Maiolino}, R., {et~al.} 2024, arXiv e-prints, arXiv:2412.04224

\bibitem[{{McClymont} {et~al.}(2025){McClymont}, {Tacchella}, {D'Eugenio}, {Witten}, {Ji}, {Smith}, {Maiolino}, {Arribas}, {Scholtz}, {Simmonds}, \& {Witstok}}]{McClymont2025}
{McClymont}, W., {Tacchella}, S., {D'Eugenio}, F., {et~al.} 2025, \mnras, 540, 190

\bibitem[{{Naidu} {et~al.}(2025){Naidu}, {Matthee}, {Katz}, {de Graaff}, {Oesch}, {Smith}, {Greene}, {Brammer}, {Weibel}, {Hviding}, {Chisholm}, {Labb\textbackslash'e}, {Simcoe}, {Witten}, {Atek}, {Baggen}, {Belli}, {Bezanson}, {Boogaard}, {Bose}, {Covelo-Paz}, {Dayal}, {Fudamoto}, {Furtak}, {Giovinazzo}, {Goulding}, {Gronke}, {Heintz}, {Hirschmann}, {Illingworth}, {Inoue}, {Johnson}, {Leja}, {Leonova}, {McConachie}, {Maseda}, {Natarajan}, {Nelson}, {Setton}, {Shivaei}, {Sobral}, {Stefanon}, {Tacchella}, {Toft}, {Torralba}, {van Dokkum}, {van der Wel}, {Volonteri}, {Walter}, {Wang}, \& {Watson}}]{Naidu2025}
{Naidu}, R.~P., {Matthee}, J., {Katz}, H., {et~al.} 2025, arXiv e-prints, arXiv:2503.16596

\bibitem[{{Nakazato} \& {Ferrara}(2024)}]{Nakazato2024}
{Nakazato}, Y. \& {Ferrara}, A. 2024, arXiv e-prints, arXiv:2412.07598

\bibitem[{{Netzer}(1975)}]{Netzer1975}
{Netzer}, H. 1975, \mnras, 171, 395

\bibitem[{Newville {et~al.}(2024)Newville, Otten, Nelson, Stensitzki, Ingargiola, Allan, Fox, Carter, Michał, Osborn, Pustakhod, Weigand, lneuhaus, Aristov, Glenn, Mark, mgunyho, Deil, Hansen, Pasquevich, Foks, Zobrist, Frost, Stuermer, Jaskula, Caldwell, Eendebak, Pompili, Nielsen, \& Persaud}]{lmfit-newville2024}
Newville, M., Otten, R., Nelson, A., {et~al.} 2024, lmfit/lmfit-py: 1.3.2

\bibitem[{{Ni} {et~al.}(2024){Ni}, {Chen}, {Zhou}, {Park}, {Yang}, {DiMatteo}, {Bird}, \& {Croft}}]{Ni2024}
{Ni}, Y., {Chen}, N., {Zhou}, Y., {et~al.} 2024, arXiv e-prints, arXiv:2409.10666

\bibitem[{{Ni} {et~al.}(2022){Ni}, {Di Matteo}, {Bird}, {Croft}, {Feng}, {Chen}, {Tremmel}, {DeGraf}, \& {Li}}]{Ni2022}
{Ni}, Y., {Di Matteo}, T., {Bird}, S., {et~al.} 2022, \mnras, 513, 670

\bibitem[{{Pollock} {et~al.}(2025){Pollock}, {Gottumukkala}, {Heintz}, {Brammer}, {Roberts-Borsani}, {Oesch}, {Witstok}, {Arellano-C{\'o}rdova}, {Cullen}, {Scholte}, {Terp}, {Rowland}, {Sneppen}, {Ito}, {Valentino}, {Matthee}, {Watson}, \& {Toft}}]{Pollock2025}
{Pollock}, C.~L., {Gottumukkala}, R., {Heintz}, K.~E., {et~al.} 2025, arXiv e-prints, arXiv:2506.15779

\bibitem[{Pérez-González {et~al.}(2024)Pérez-González, Barro, Rieke, Lyu, Rieke, Alberts, Williams, Hainline, Sun, Puskás, Annunziatella, Baker, Bunker, Egami, Ji, Johnson, Robertson, Rodríguez Del~Pino, Rujopakarn, Shivaei, Tacchella, Willmer, \& Willott}]{PerezGonzalez2024}
Pérez-González, P.~G., Barro, G., Rieke, G.~H., {et~al.} 2024, The Astrophysical Journal, 968, 4

\bibitem[{{Rinaldi} {et~al.}(2024){Rinaldi}, {Bonaventura}, {Rieke}, {Alberts}, {Caputi}, {Baker}, {Baum}, {Bhatawdekar}, {Bunker}, {Carniani}, {Curtis-Lake}, {D'Eugenio}, {Egami}, {Ji}, {Hainline}, {Helton}, {Lin}, {Lyu}, {Johnson}, {Ma}, {Maiolino}, {P{\'e}rez-Gonz{\'a}lez}, {Rieke}, {Robertson}, {Shivaei}, {Stone}, {Sun}, {Tacchella}, {{\"U}bler}, {Williams}, {Willmer}, {Willott}, {Zhang}, \& {Zhu}}]{Rinaldi2024}
{Rinaldi}, P., {Bonaventura}, N., {Rieke}, G.~H., {et~al.} 2024, arXiv e-prints, arXiv:2411.14383

\bibitem[{{Rusakov} {et~al.}(2025){Rusakov}, {Watson}, {Nikopoulos}, {Brammer}, {Gottumukkala}, {Harvey}, {Heintz}, {Nielsen}, {Sim}, {Sneppen}, {Vijayan}, {Adams}, {Austin}, {Conselice}, {Goolsby}, {Toft}, \& {Witstok}}]{Rusakov2025}
{Rusakov}, V., {Watson}, D., {Nikopoulos}, G.~P., {et~al.} 2025, arXiv e-prints, arXiv:2503.16595

\bibitem[{{Scarlata} {et~al.}(2024){Scarlata}, {Hayes}, {Panagia}, {Mehta}, {Haardt}, \& {Bagley}}]{Scarlata2024}
{Scarlata}, C., {Hayes}, M., {Panagia}, N., {et~al.} 2024, arXiv e-prints, arXiv:2404.09015

\bibitem[{{Setton} {et~al.}(2024){Setton}, {Greene}, {de Graaff}, {Ma}, {Leja}, {Matthee}, {Bezanson}, {Boogaard}, {Cleri}, {Katz}, {Labbe}, {Maseda}, {McConachie}, {Miller}, {Price}, {Suess}, {van Dokkum}, {Wang}, {Weibel}, {Whitaker}, \& {Williams}}]{Setton2024}
{Setton}, D.~J., {Greene}, J.~E., {de Graaff}, A., {et~al.} 2024, arXiv e-prints, arXiv:2411.03424

\bibitem[{{Setton} {et~al.}(2025){Setton}, {Greene}, {Spilker}, {Williams}, {Labbe}, {Ma}, {Wang}, {Whitaker}, {Leja}, {de Graaff}, {Alberts}, {Bezanson}, {Boogaard}, {Brammer}, {Cutler}, {Cleri}, {Cooper}, {Dayal}, {Fujimoto}, {Furtak}, {Goulding}, {Hirschmann}, {Kokorev}, {Maseda}, {McConachie}, {Matthee}, {Miller}, {Naidu}, {Oesch}, {Pan}, {Price}, {Suess}, {Weaver}, {Xiao}, {Zhang}, \& {Zitrin}}]{Setton2025}
{Setton}, D.~J., {Greene}, J.~E., {Spilker}, J.~S., {et~al.} 2025, arXiv e-prints, arXiv:2503.02059

\bibitem[{{Skilling}(2004)}]{Skilling2004}
{Skilling}, J. 2004, in American Institute of Physics Conference Series, Vol. 735, Bayesian Inference and Maximum Entropy Methods in Science and Engineering: 24th International Workshop on Bayesian Inference and Maximum Entropy Methods in Science and Engineering, ed. R.~{Fischer}, R.~{Preuss}, \& U.~V. {Toussaint} (AIP), 395--405

\bibitem[{Skilling(2006)}]{Skilling2006}
Skilling, J. 2006, Bayesian Analysis, 1, 833

\bibitem[{Sneppen {et~al.}(2023)Sneppen, Watson, Bauswein, Just, Kotak, Nakar, Poznanski, \& Sim}]{Sneppen2023}
Sneppen, A., Watson, D., Bauswein, A., {et~al.} 2023, Nature, 614, 436

\bibitem[{{Speagle}(2020)}]{Speagle2020}
{Speagle}, J.~S. 2020, \mnras, 493, 3132

\bibitem[{{Sun} \& {Yan}(2025)}]{Sun2025-flatdecs}
{Sun}, B. \& {Yan}, H. 2025, arXiv e-prints, arXiv:2503.21896

\bibitem[{Taylor {et~al.}(2025)Taylor, Kokorev, Kocevski, Akins, Cullen, Dickinson, Finkelstein, Haro, Bromm, Giavalisco, Inayoshi, Juneau, Leung, Perez-Gonzalez, Somerville, Trump, Amorin, Barro, Burgarella, Brooks, Carnall, Casey, Cheng, Chisholm, Chworowsky, Davis, Donnan, Dunlop, Ellis, Fernandez, Fujimoto, Grogin, Gupta, Hathi, Jung, Hirschmann, Kartaltepe, Koekemoer, Larson, Leung, Llerena, Lucas, McLeod, McLure, Napolitano, Papovich, Stanton, Tripodi, Wang, Wilkins, Yung, \& Zavala}]{Taylor2025}
Taylor, A.~J., Kokorev, V., Kocevski, D.~D., {et~al.} 2025, CAPERS-LRD-z9: A Gas Enshrouded Little Red Dot Hosting a Broad-line AGN at z=9.288

\bibitem[{{Topping} {et~al.}(2024){Topping}, {Stark}, {Senchyna}, {Plat}, {Zitrin}, {Endsley}, {Charlot}, {Furtak}, {Maseda}, {Smit}, {Mainali}, {Chevallard}, {Molyneux}, \& {Rigby}}]{Topping2024}
{Topping}, M.~W., {Stark}, D.~P., {Senchyna}, P., {et~al.} 2024, \mnras, 529, 3301

\bibitem[{{Torralba} {et~al.}(2025){Torralba}, {Matthee}, {Pezzulli}, {Naidu}, {Ishikawa}, {Brammer}, {Chang}, {Chisholm}, {de Graaff}, {D'Eugenio}, {Di Cesare}, {Eilers}, {Greene}, {Gronke}, {Iani}, {Kokorev}, {Kotiwale}, {Kramarenko}, {Ma}, {Mascia}, {Navarrete}, {Nelson}, {Oesch}, {Simcoe}, \& {Wuyts}}]{Torralba2025}
{Torralba}, A., {Matthee}, J., {Pezzulli}, G., {et~al.} 2025, arXiv e-prints, arXiv:2510.00103

\bibitem[{{Trebitsch} {et~al.}(2021){Trebitsch}, {Dubois}, {Volonteri}, {Pfister}, {Cadiou}, {Katz}, {Rosdahl}, {Kimm}, {Pichon}, {Beckmann}, {Devriendt}, \& {Slyz}}]{Trebitsch2025}
{Trebitsch}, M., {Dubois}, Y., {Volonteri}, M., {et~al.} 2021, \aap, 653, A154

\bibitem[{{Trinca} {et~al.}(2023){Trinca}, {Schneider}, {Maiolino}, {Valiante}, {Graziani}, \& {Volonteri}}]{Trinca2023}
{Trinca}, A., {Schneider}, R., {Maiolino}, R., {et~al.} 2023, \mnras, 519, 4753

\bibitem[{{Volonteri} {et~al.}(2025){Volonteri}, {Trebitsch}, {Greene}, {Dubois}, {Dong-Paez}, {Habouzit}, {Lupi}, {Ma}, {Beckmann}, {Dayal}, \& {Schneider}}]{Volonteri2025}
{Volonteri}, M., {Trebitsch}, M., {Greene}, J.~E., {et~al.} 2025, \aap, 695, A33

\bibitem[{Wang {et~al.}(2024)Wang, Leja, de~Graaff, Brammer, Weibel, van Dokkum, Baggen, Suess, Greene, Bezanson, Cleri, Hirschmann, Labbé, Matthee, McConachie, Naidu, Nelson, Oesch, Setton, \& Williams}]{Wang2024b}
Wang, B., Leja, J., de~Graaff, A., {et~al.} 2024, The Astrophysical Journal Letters, 969, L13

\bibitem[{Williams {et~al.}(2024)Williams, Alberts, Ji, Hainline, Lyu, Rieke, Endsley, Suess, Sun, Johnson, Florian, Shivaei, Rujopakarn, Baker, Bhatawdekar, Boyett, Bunker, Cameron, Carniani, Charlot, Curtis-Lake, DeCoursey, de~Graaff, Egami, Eisenstein, Gibson, Hausen, Helton, Maiolino, Maseda, Nelson, Pérez-González, Rieke, Robertson, Saxena, Tacchella, Willmer, \& Willott}]{Williams2024}
Williams, C.~C., Alberts, S., Ji, Z., {et~al.} 2024, The Astrophysical Journal, 968, 34

\bibitem[{Yue {et~al.}(2024)Yue, Eilers, Ananna, Panagiotou, Kara, \& Miyaji}]{Yue2024}
Yue, M., Eilers, A.-C., Ananna, T.~T., {et~al.} 2024, The Astrophysical Journal Letters, 974, L26

\bibitem[{{Zhang} {et~al.}(2025){Zhang}, {Jiang}, {Liu}, {Ho}, \& {Inayoshi}}]{Zhang2025-NarrowLRDS}
{Zhang}, Z., {Jiang}, L., {Liu}, W., {Ho}, L.~C., \& {Inayoshi}, K. 2025, arXiv e-prints, arXiv:2506.04350

\end{thebibliography}

%
\onecolumn
\begin{appendix}
\section{Object SEDs}
\begin{figure*}[h!]       
  \centering
    \includegraphics[scale = 0.6]{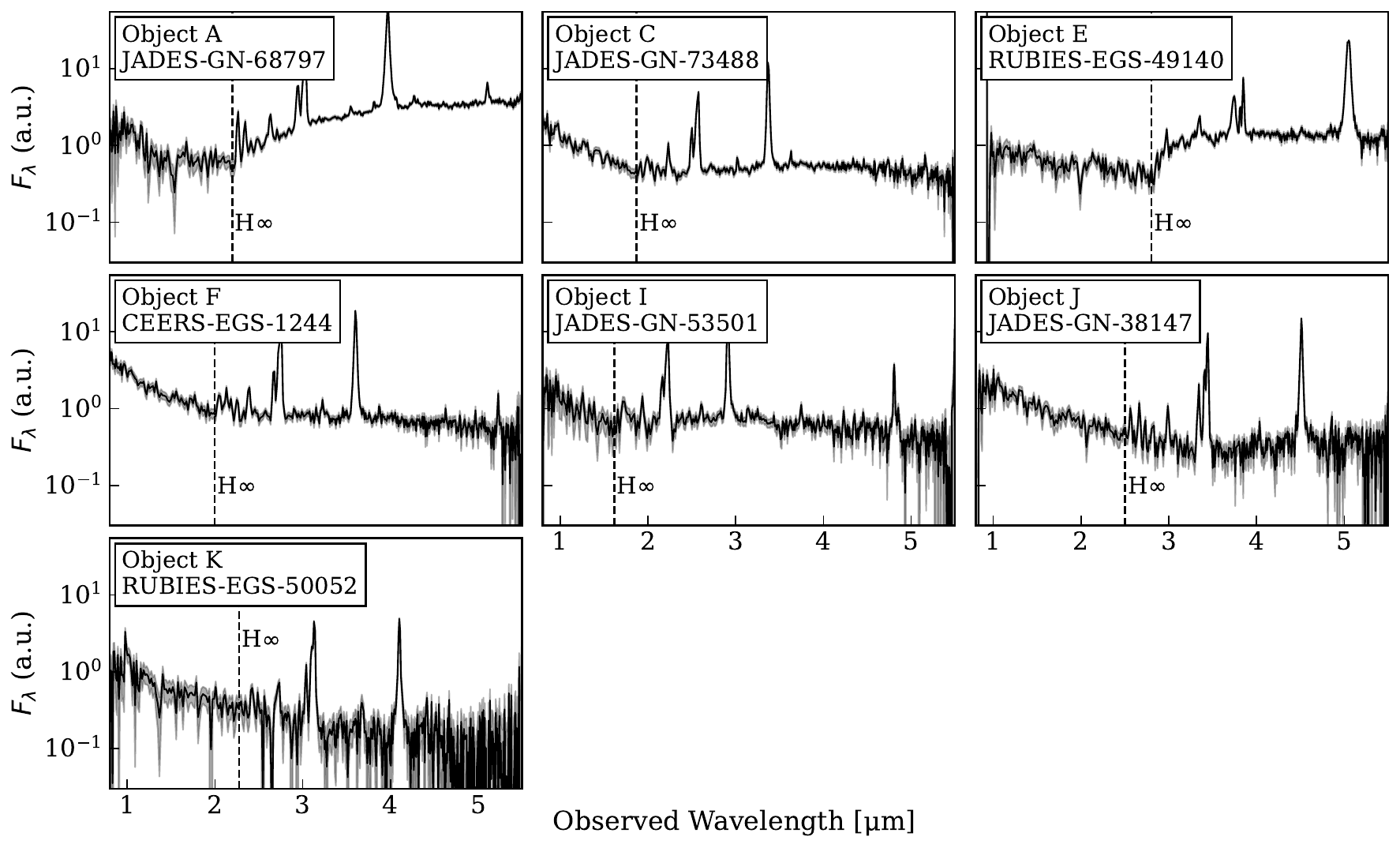}
    \caption{Spectral energy distribution for LRDs in the sample.}
    \label{fig:appendix-SEDs}

\end{figure*}
\section{Line Fluxes}
\renewcommand{\arraystretch}{1.3}
\begin{table}[h]
\caption{Flux estimates for the narrow and broad components of \ha, \hb, \hg, and \hd lines of the objects investigated in this work, in units of $10^{-19}$\,erg\,s$^{-1}$\,cm$^{-2}$. Object~A (JADES-GN-68797) has no detectable narrow component, while there is no spectral coverage of \hd for Object~J (JADES-GN-38147)}
\label{tab:fluxes}
\begin{tabular}{@{}cccccccccc@{}}
\cmidrule(l){3-10}
\multicolumn{2}{c}{}    & \multicolumn{2}{c}{\ha}                                  & \multicolumn{2}{c}{\hb}                                  & \multicolumn{2}{c}{\hg}                            & \multicolumn{2}{c}{\hd}             \\ \cmidrule(l){3-10} 
ID & Survey-Field-MSAID & Narrow                    & Broad                       & Narrow                     & Broad                      & Narrow                  & Broad                   & Narrow                  & Broad    \\ \midrule
A  & JADES-GN-68797     & -                         & $2183.19^{+5.02}_{-27.06}$  & -                          & $131.73^{+13.43}_{-12.56}$ & -                       & $36.26^{+6.52}_{-6.55}$ & -                       & $<25.73$ \\
C  & JADES-GN-73488     & $34.93^{+2.52}_{-2.46}$   & $306.35^{+6.08}_{-4.74}$    & $18.591^{+0.839}_{-0.853}$ & $15.60^{+2.38}_{-2.10}$    & $7.44^{+1.39}_{-1.48}$  & $<17.95$                & $5.34^{+1.26}_{-1.20}$  & $<15.79$ \\
E  & RUBIES-EGS-49140   & $23.23^{+10.59}_{-3.48}$  & $1202.75^{+14.02}_{-25.10}$ & $7.62^{+1.64}_{-1.66}$     & $137.89^{+5.17}_{-5.96}$   & $4.81^{+0.79}_{-1.07}$  & $32.36^{+3.30}_{-5.16}$ & $<2.38$                 & $<7.79$  \\
F  & CEERS-EGS-1244     & $77.83^{+5.44}_{-15.71}$  & $486.96^{+16.94}_{-7.09}$   & $42.21^{+4.20}_{-3.72}$    & $<71.98$                   & $12.67^{+3.08}_{-2.88}$ & $<41.21$                & $11.21^{+3.38}_{-2.61}$ & $<15.98$ \\
I  & JADES-GN-53501     & $53.37^{+12.64}_{-12.82}$ & $334.70^{+16.88}_{-17.25}$  & $13.89^{+2.80}_{-2.71}$    & $39.68^{+6.06}_{-5.53}$    & $<11.38$                & $<21.12$                & $<4.24$                 & $<29.00$ \\
J  & JADES-GN-38147     & $88.24^{+11.90}_{-11.77}$ & $283.55^{+12.56}_{-13.20}$  & $30.33^{+2.14}_{-1.86}$    & $<10.97$                   & $14.36^{+2.28}_{-2.01}$ & $<1.65$                 & -                       & -        \\
K  & RUBIES-EGS-50052   & $31.20^{+7.20}_{-6.23}$   & $60.09^{+6.52}_{-8.32}$     & $9.31^{+1.96}_{-2.09}$     & $18.26^{+5.92}_{-5.82}$    & $<7.49$                 & $<6.72$                 & $<3.98$                 & $<7.07$  \\ \bottomrule
\end{tabular}
\end{table}

\section{Fits}

In this section we provide the fit results for all the lines examined in this work. 

\begin{figure*}         
  \centering
  \begin{subfigure}{.49\textwidth}
    \includegraphics[width=80mm]{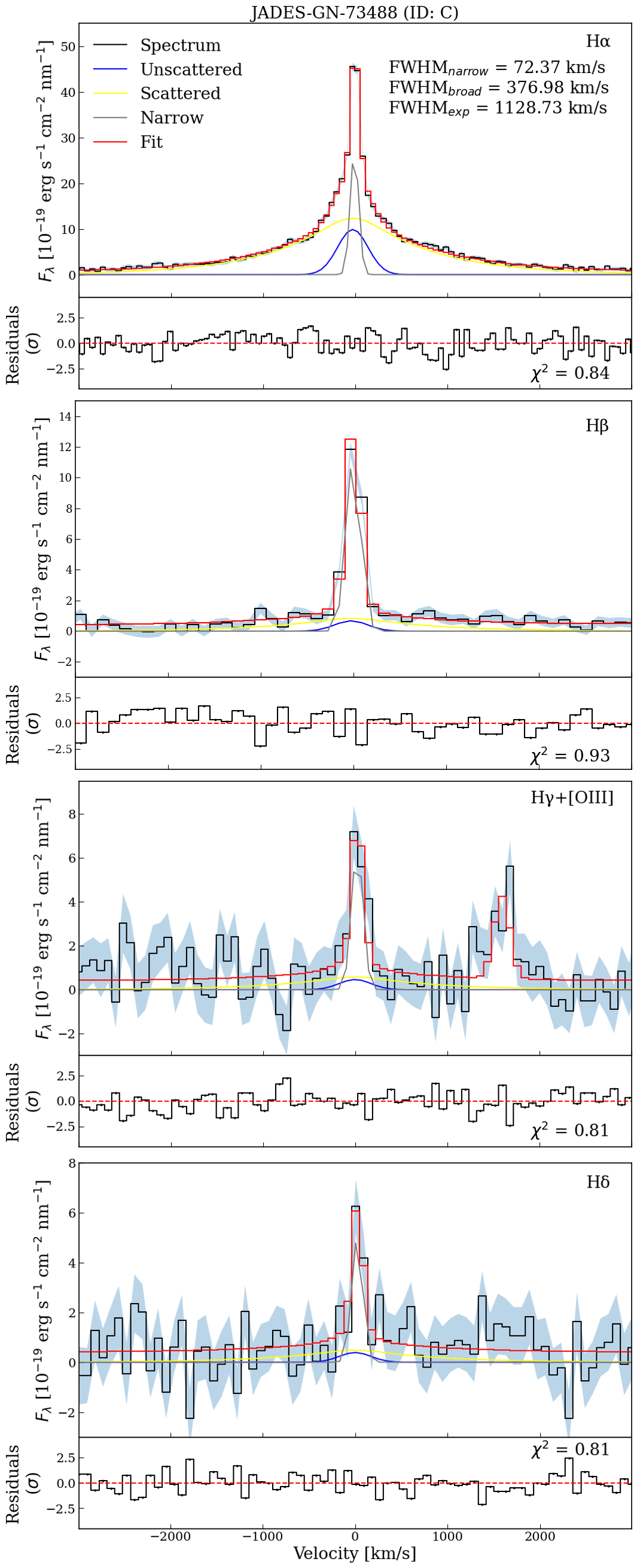}
    \caption{}
    \label{fig:objectC}
  \end{subfigure}\hfill
  \begin{subfigure}{.49\textwidth}
    \includegraphics[width=80mm]{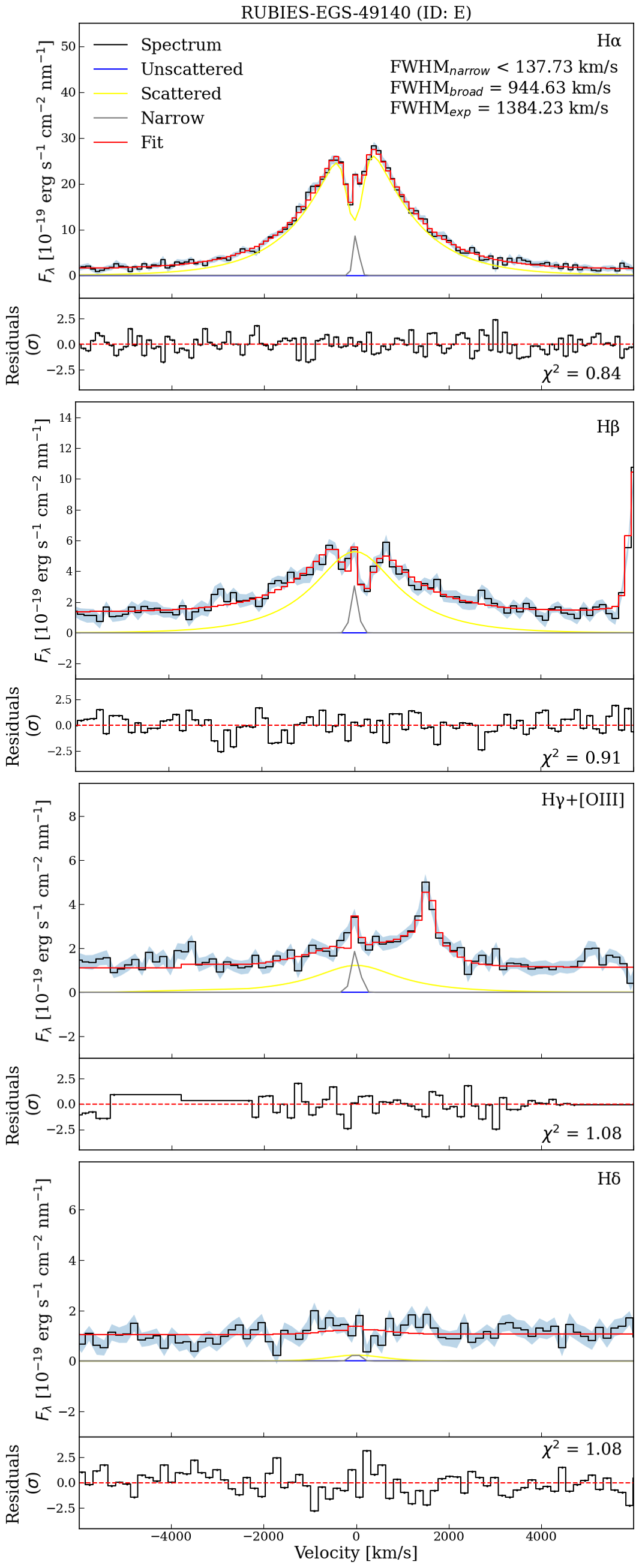}%
    \caption{}
    \label{fig:objectE}
  \end{subfigure}
    \caption{Fits to the Balmer lines of Objects~C (\emph{left}, JADES-GN-73488) and E (\emph{right} RUBIES-EGS-49140).}
  \label{fig:appendix-c+e}

\end{figure*}
\begin{figure*}          
  \centering
  \begin{subfigure}{.49\textwidth}
    \includegraphics[width=80mm]{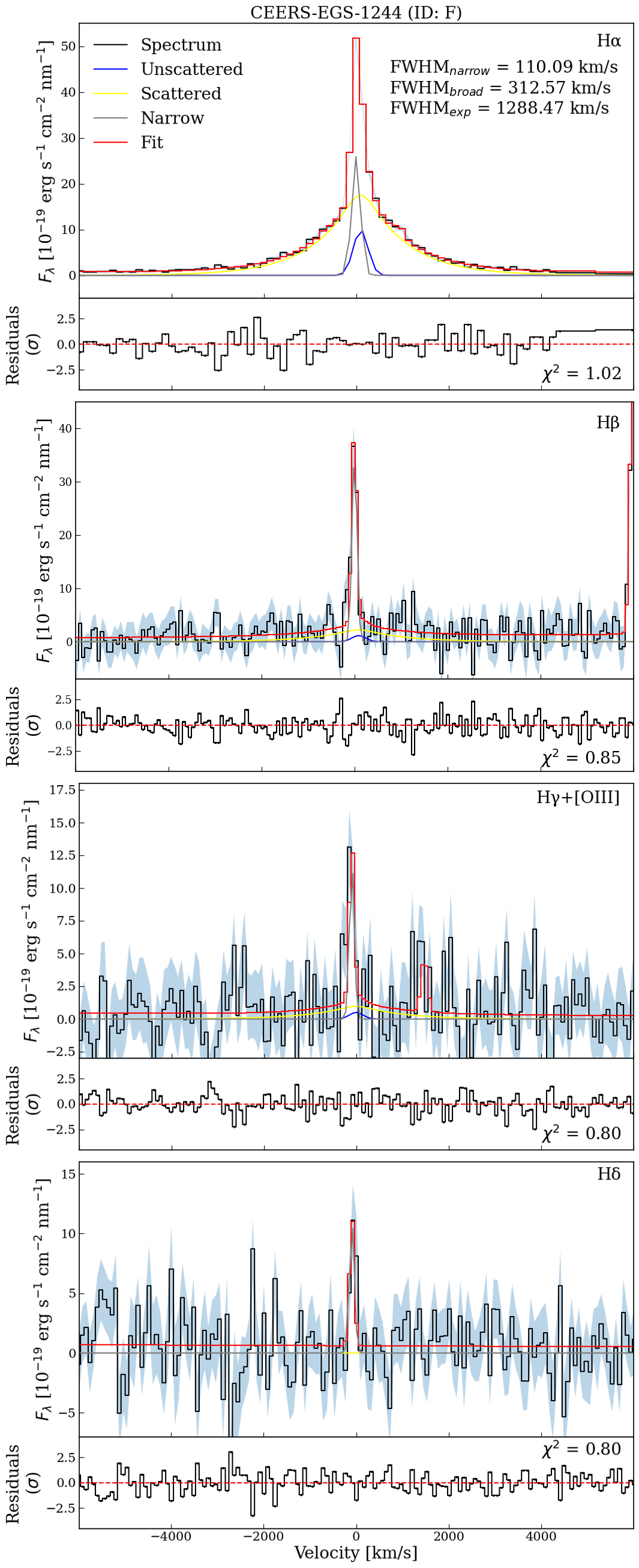}
    \caption{}
    \label{fig:objectF}
  \end{subfigure}\hfill
  \begin{subfigure}{.49\textwidth}
    \includegraphics[width=80mm]{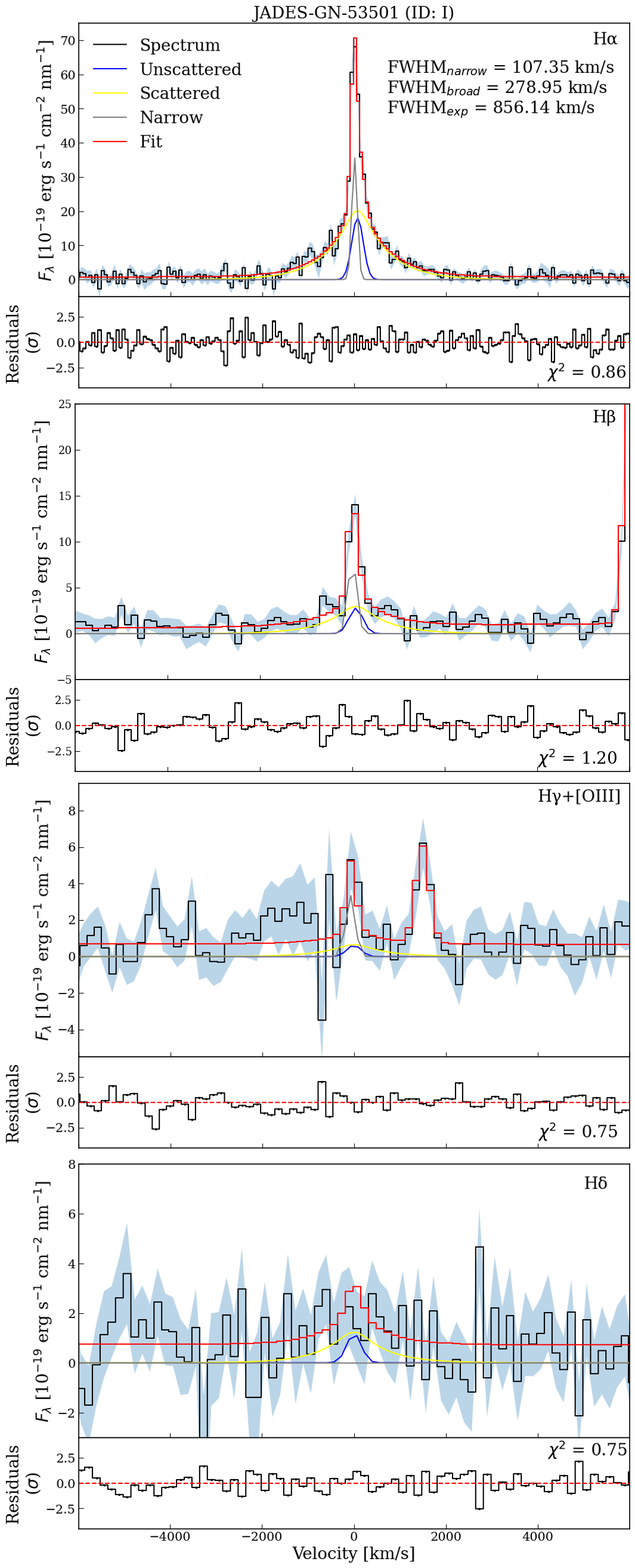}%
    \caption{}
    \label{fig:objectI}
  \end{subfigure}
    \caption{Fits to the Balmer lines of Objects~F (\emph{left}, CEERS-EGS-1244) and I (\emph{right} JADES-GN-53501).}
  \label{fig:appendix-f+i}

\end{figure*}
\begin{figure*}            
  \centering
  \begin{subfigure}{.49\textwidth}
    \includegraphics[width=80mm]{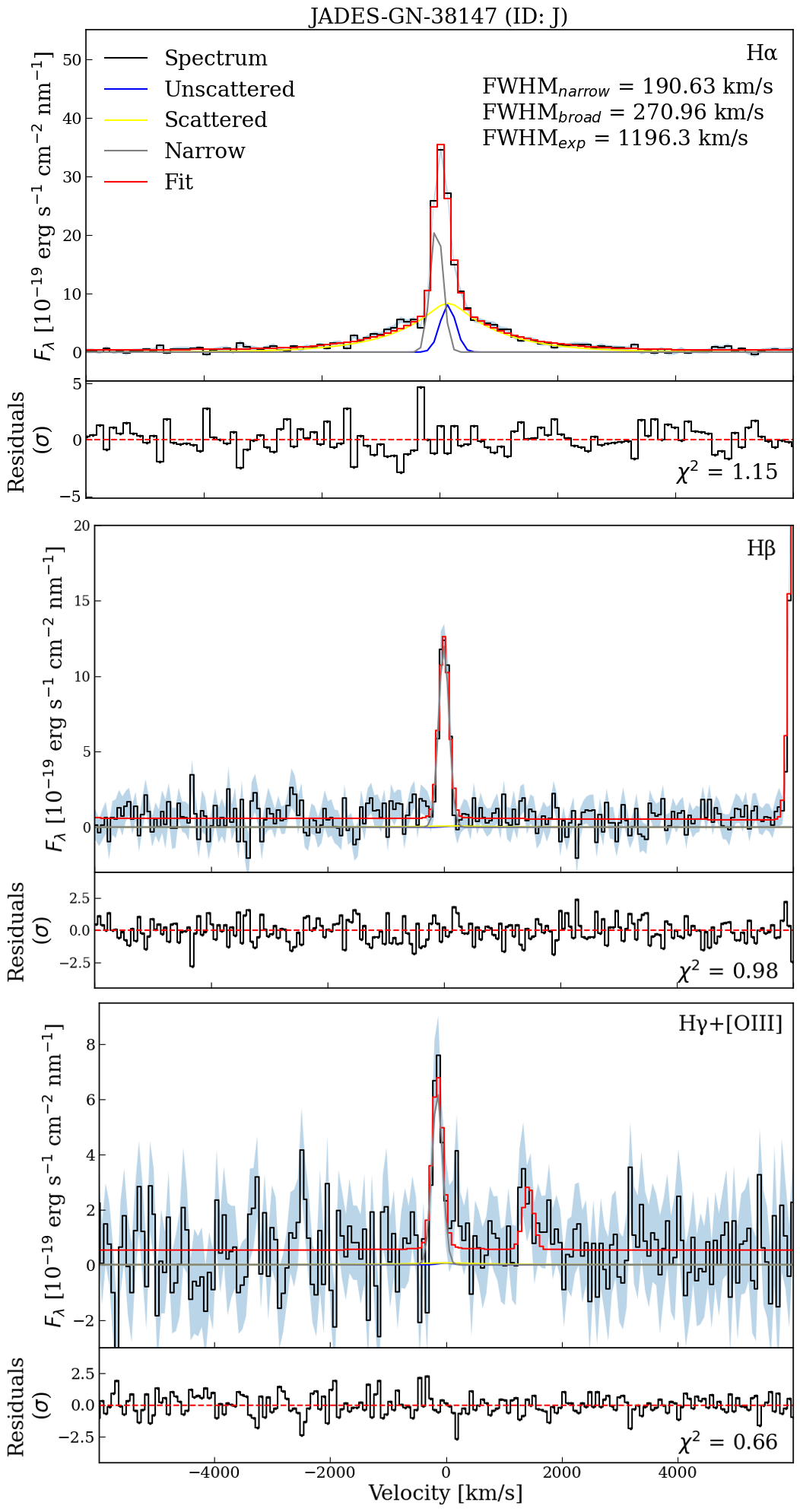}
    \caption{}
    \label{fig:objectJ}
  \end{subfigure}\hfill
  \begin{subfigure}{.49\textwidth}
    \includegraphics[width=80mm]{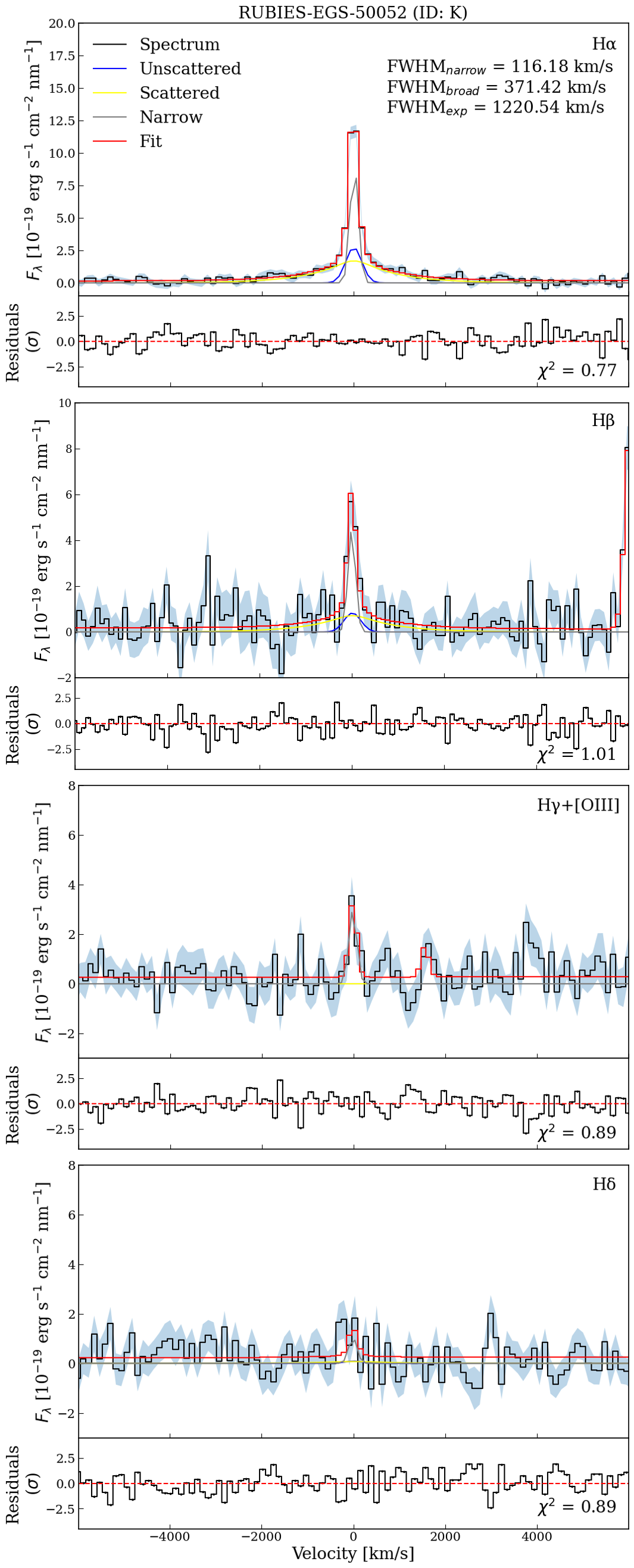}%
    \caption{}
    \label{fig:objectK}
  \end{subfigure}
  \label{fig:appendix-j+k}
    \caption{Fits to the Balmer lines of Objects~J (\emph{left}, JADES-GN-38147) and K (\emph{right} RUBIES-EGS-50052).}
\end{figure*}

\section{Posteriors}
We provide the distribution of the relevant posterior parameter values for the Fiducial models for all fitted lines per object. The parameters presented for the narrow H$\alpha_{\rm{nr}}$, unscattered H$\alpha_{\rm{un}}$, scattered H$\alpha_{exp}$ and absorbed H$\alpha_{\rm{abs}}$ include: the width $\sigma$ in [nm] and the amplitude $A$. The PCygni absorption/emission feature is modelled using the maximum ejecta velocity v$_{\rm{out}}$, the velocity of the photosphere v$_{\rm{phot}}$ in units of the speed of light $c$, as well as the optical depth of the absorption, $\tau$, using code presented in \cite{Sneppen2023}, originally adapted from Ulrich Noebauer’s code \href{https://github.com/unoebauer/public-astro-tools}{https://github.com/unoebauer/public-astro-tools}. The components of the profiles of \hb, \hg and \hd are scaled to the same components of the respective \ha lines with a scaling factor, so that e.g. H$\beta_{\rm{nr}}$ = $A_{\rm{nr}}$H$\alpha_{\rm{nr}}$. We divide the lines of each object into narrow and broad, so that the unscattered and scattered components are scaled with the same factor in all higher order Balmer lines. Object~A and E are an exception; for Object~A there is no narrow emission detected, while for Object~E no light emerges without being scattered. All corner plots are made using \texttt{corner.py} \citep{corner} 

\begin{figure}
    \centering
    \includegraphics[width=0.8\linewidth]{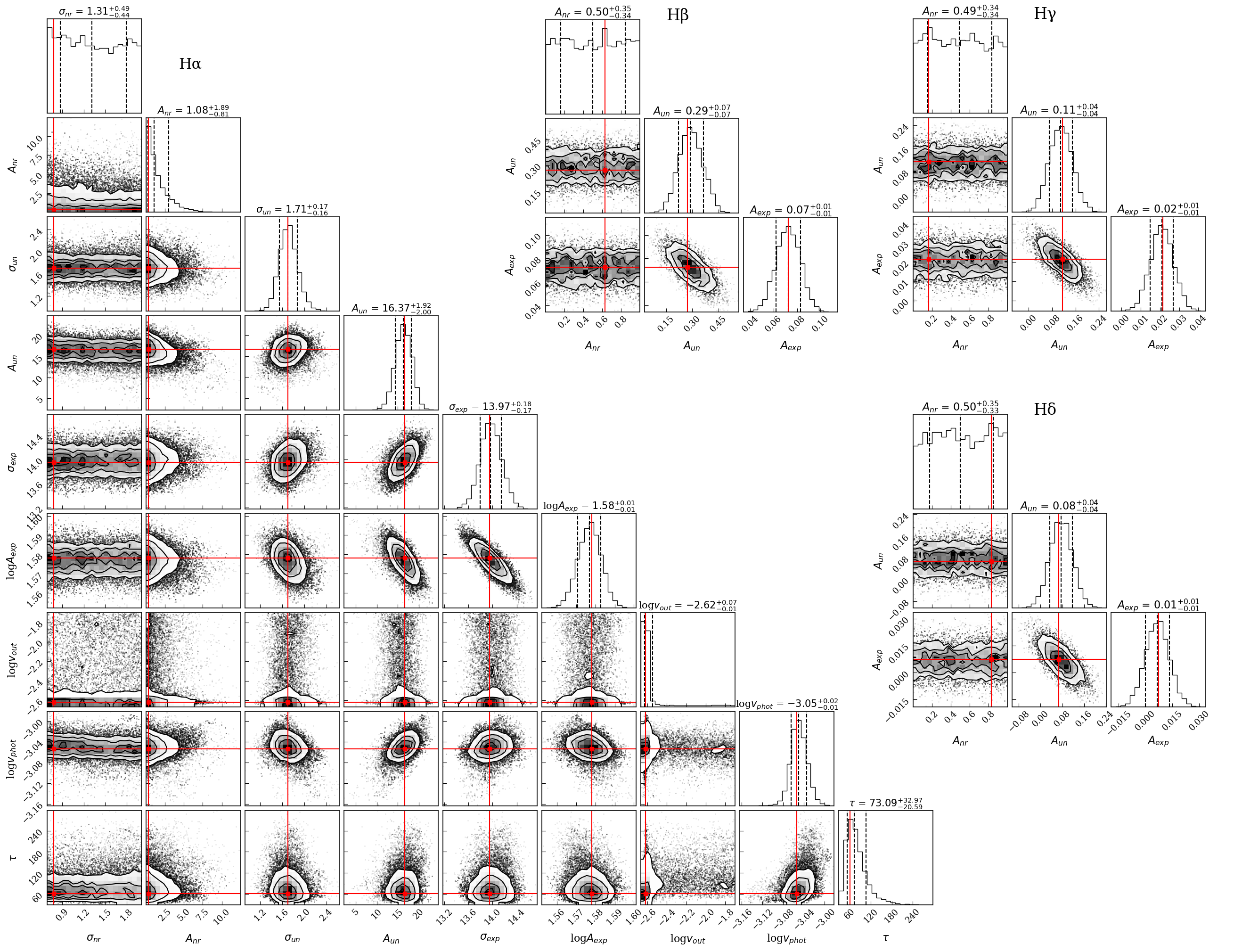}
    \caption{Posterior corner plot for JADES-GN-68797 (Object~A). All lines denoted in their respective corner plots.}
    \label{fig:posteriors-a}
\end{figure}

\begin{figure}
    \centering
    \includegraphics[width=0.8\linewidth]{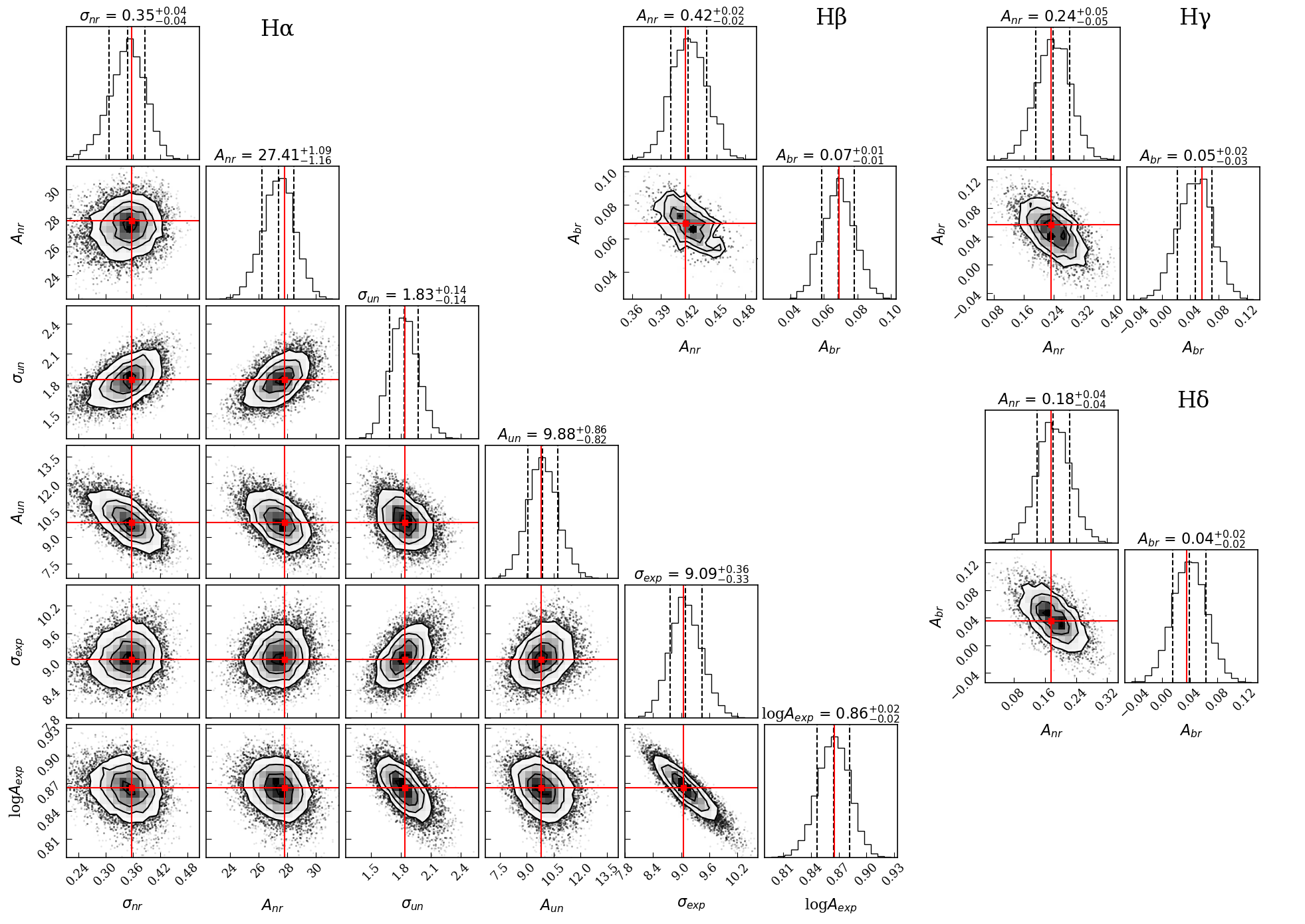}
    \caption{Posterior probability corner plot for Object~C (JADES-GN-73488). The relevant Balmer lines are denoted on their respective corner plots.}\label{fig:posteriors-c}
\end{figure}

\begin{figure}
    \centering
    \includegraphics[width=0.8\linewidth]{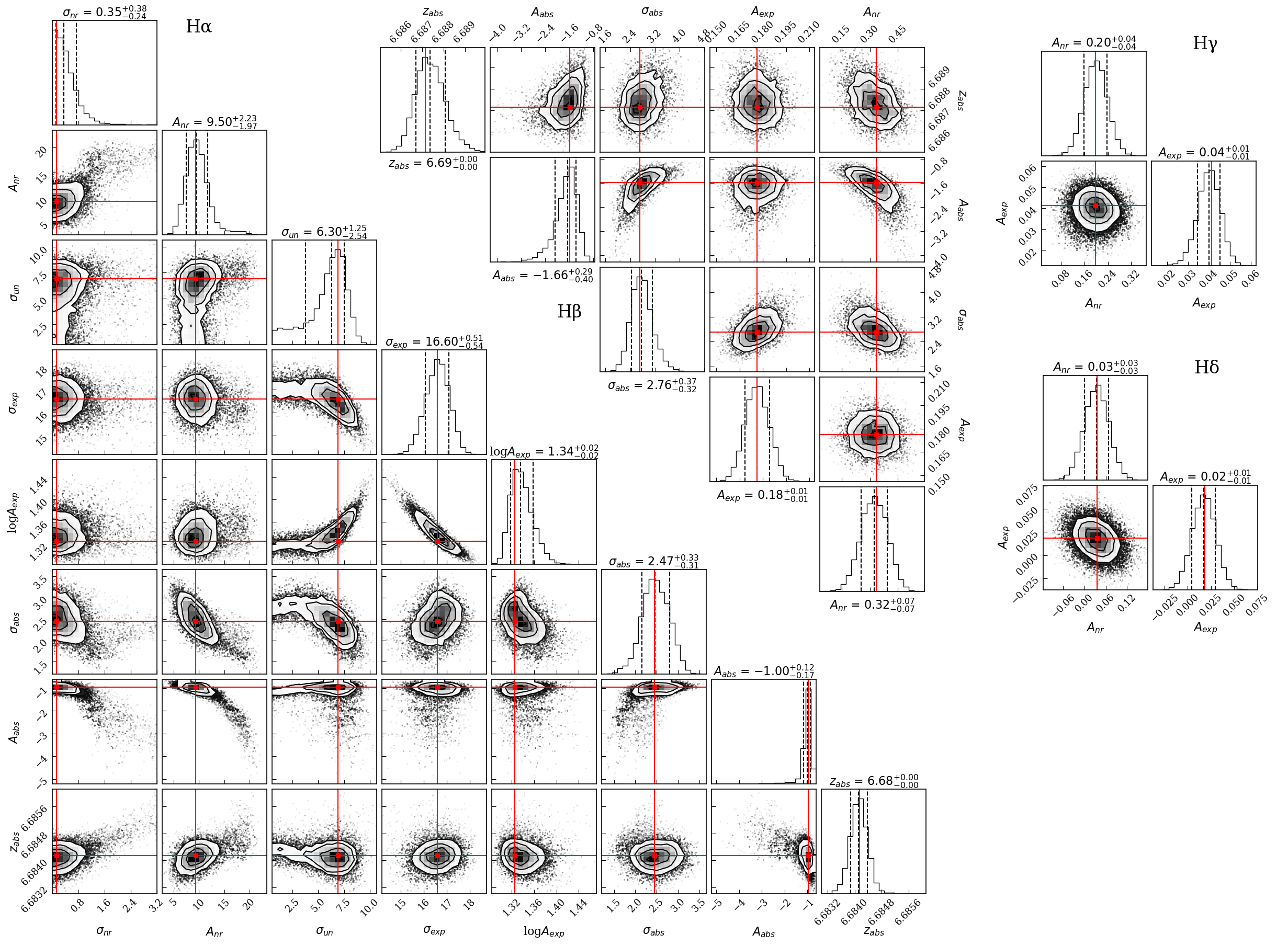}
    \caption{As Fig.~\ref{fig:posteriors-c} but for Object~E (RUBIES-4233-49140).}
    \label{fig:posteriors-e}
\end{figure}

\begin{figure}
    \centering
    \includegraphics[width=0.8\linewidth]{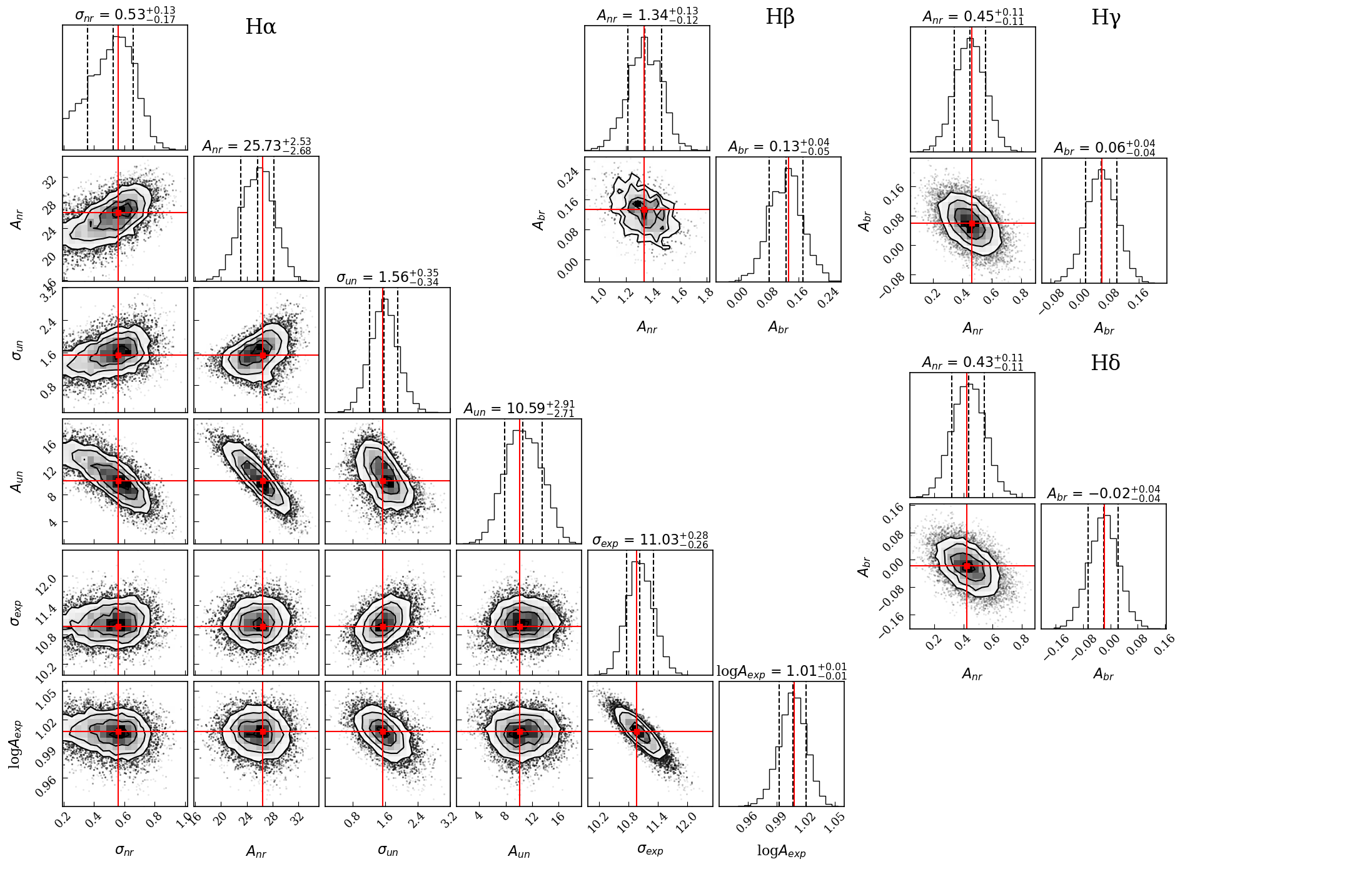}
    \caption{As Fig.~\ref{fig:posteriors-c} but for Object~F (CEERS-EGS-1244).}
    \label{fig:posteriors-f}
\end{figure}

\begin{figure}
    \centering
    \includegraphics[width=0.8\linewidth]{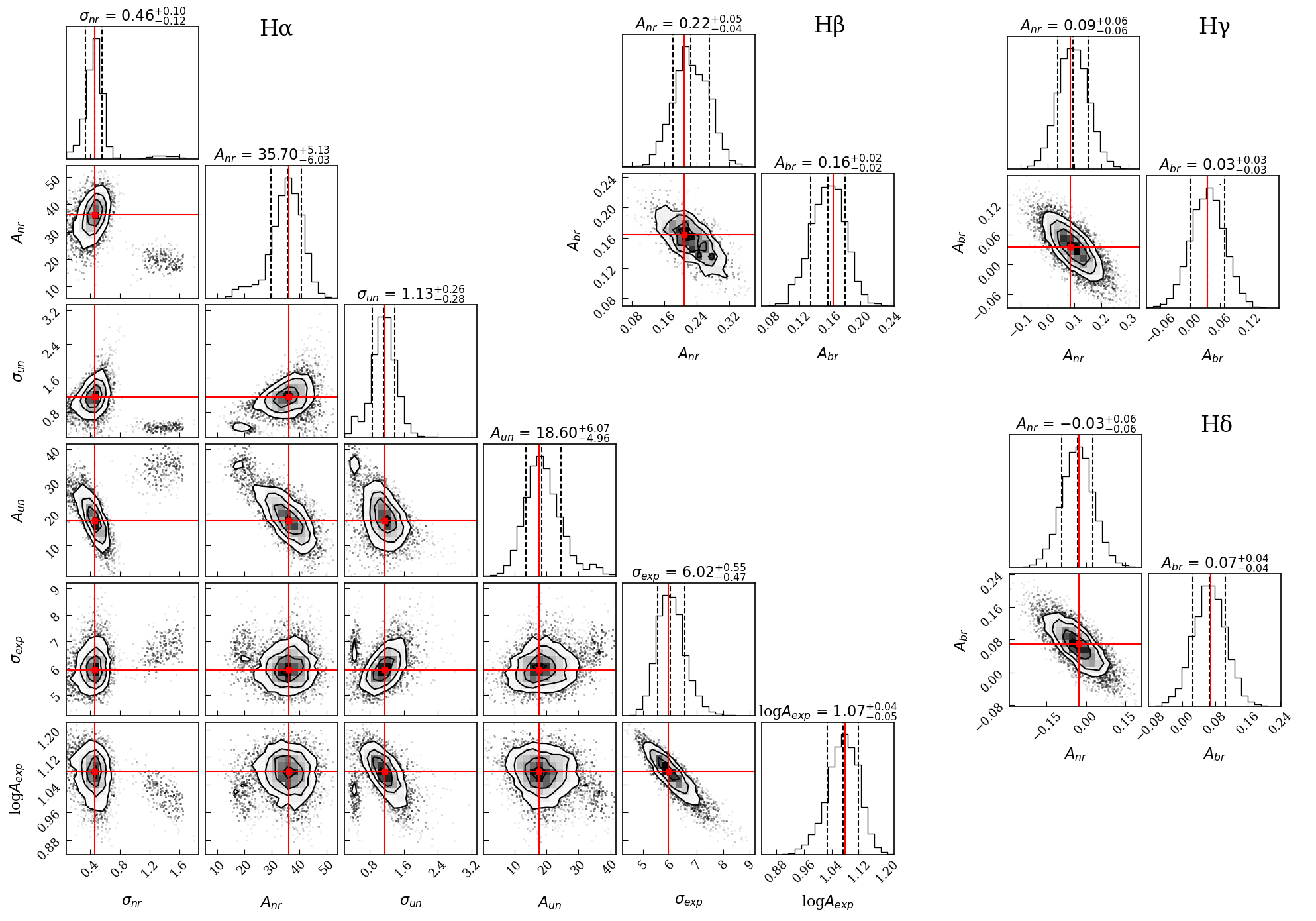}
    \caption{As Fig.~\ref{fig:posteriors-c} but for Object~I (JADES-GN-53501).}
    \label{fig:posteriors-i}
\end{figure}

\begin{figure}
    \centering
    \includegraphics[width=0.8\linewidth]{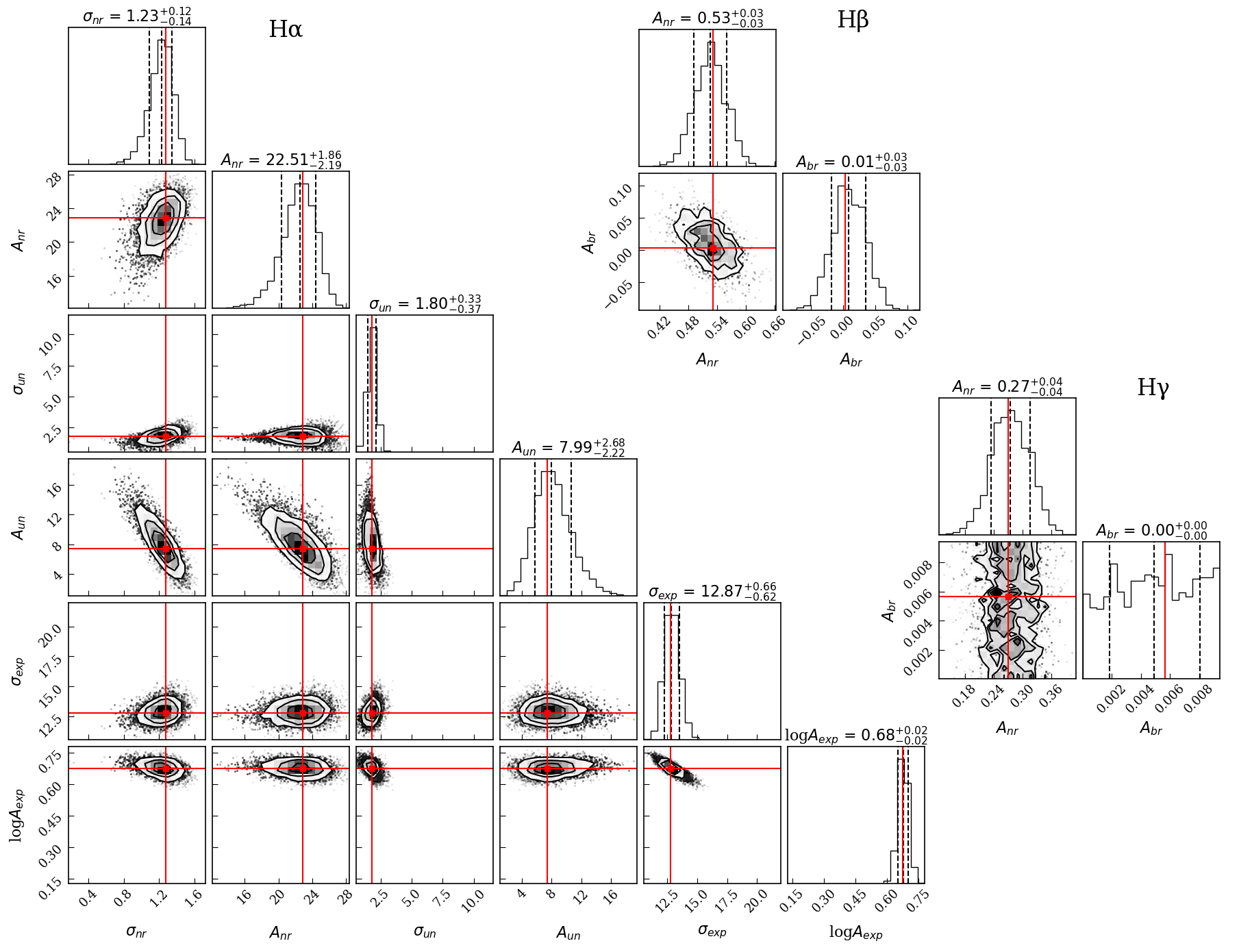}
    \caption{As Fig.~\ref{fig:posteriors-c} but for Object~J (JADES-GN-38147).}
    \label{fig:posteriors-j}
\end{figure}

\begin{figure}
    \centering
    \includegraphics[width=0.8\linewidth]{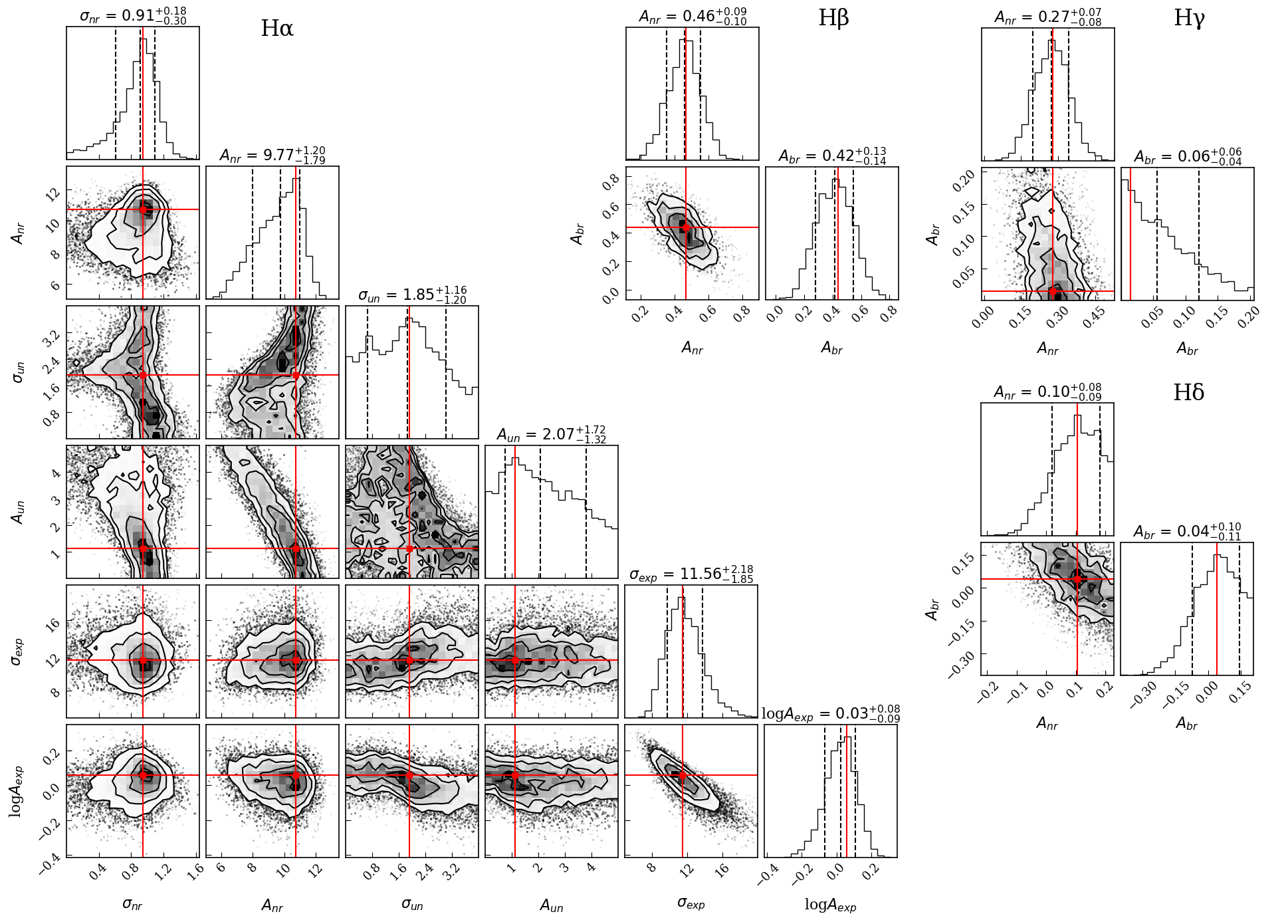}
    \caption{As Fig.~\ref{fig:posteriors-c} but for Object~K (RUBIES-EGS-50052).}
    \label{fig:posteriors-k}
\end{figure}

\end{appendix}
\end{document}